\documentclass[aps,prx,twocolumn,english,superscriptaddress]{revtex4}
\usepackage[utf8]{inputenc}

\usepackage{ulem} 

\usepackage{placeins} 

\usepackage{amsmath}
\usepackage{amsfonts}
\usepackage{mathtools}

\usepackage{color}
\usepackage{graphicx} 
\usepackage{hyperref} 
\definecolor{myDarkGreen}{rgb}{0, 0.7, 0}

\mathchardef\mhyphen="2D

\begin{document}

\title{Understanding domain-wall encoding theoretically and experimentally}

\author{Jesse Berwald}
\affiliation{Quantum Computing Inc.~Leesburg, VA, USA}
\author{Nicholas Chancellor}
\email{nicholas.chancellor@gmail.com}
\affiliation{Quantum Computing Inc.~Leesburg, VA, USA}
\affiliation{ Physics department, Durham University Physics department Quantum Light and Matter section and Durham-Newcastle Joint Quantum Centre, South Road, Durham UK}
\author{Raouf Dridi}
\affiliation{Quantum Computing Inc.~Leesburg, VA, USA}

\begin{abstract}  
We analyze the method of encoding pairwise interactions of higher-than-binary discrete variables (these models are sometimes referred to as discrete quadratic models) into binary variables based on domain walls on one dimensional Ising chains. We discuss how this is relevant to quantum annealing, but also many gate model algorithms such as VQE and QAOA. We theoretically show that for problems of practical interest for quantum computing and assuming only quadratic interactions are available between the binary variables, it is not possible to have a more efficient general encoding in terms of number of binary variables per discrete variable. We furthermore use a D-Wave 
Advantage 1.1 flux qubit quantum annealing computer
to show that the dynamics effectively freeze later for a domain-wall encoding compared to a traditional one-hot encoding. This second result could help explain the dramatic performance improvement of domain wall over one hot which has been seen in a recent experiment on D-Wave hardware. This is an important result because usually problem encoding and the underlying physics are considered separately, our work suggests that considering them together may be a more useful paradigm. We argue that this experimental result is also likely to carry over to a number of other settings, we discuss how this has implications for gate-model and quantum-inspired algorithms.

\end{abstract}

\maketitle

\date{\today}

\medskip

\medskip

\medskip


\medskip

\medskip

\medskip

\section{Introduction}
Quantum computing shows great promise for combinatorial optimisation problems, and many proof-of-concept experiments have been performed demonstrating the potential in a variety of areas including vehicle scheduling~\cite{crispin13a}, traffic flow optimisation \cite{neukart2017,yarkoni2020}, hydrology \cite{omalley18a}, computational biology~\cite{perdomo-ortiz12a,Mulligan20}, community detection \cite{Negre20a}, graph theoretical problems \cite{Ushijima-Mwesigwa17a,Tabi20a,chen21a}, and supply chain logistics \cite{Ding21a}. While this is an area with great promise, available devices exist in relatively early stages of  development, which is often termed the noisy intermediate-scale quantum (NISQ) \cite{Preskill18a} era of quantum computing. In these early stages, it is crucial to be able to get the most out of these devices through, among other things, optimal encoding of problems \cite{Callison2022perspective}. 

Currently there are two major paradigms of quantum computing, analog quantum computing, typified by quantum annealing computers, and employing continuous time evolution, and digital gate model quantum computing which performs a series of discrete ``gate'' operations. In both settings, it is crucial to use the hardware optimally, including optimal encoding of problems, but the constraints may be different. In particular, the physical interactions underlying the operation of these devices typically do not involve more than two qubits, so it is natural to consider optimisation problems which can be expressed as a quadratic problem, those which only involve single Boolean variable terms 
and interactions between pairs of Boolean variables. One example of particular interest here are quadratic unconstrained binary optimisation problems (QUBOs). It is always possible to map an optimisation problem involving higher order interactions into a QUBO, but this will come at a cost in terms of ``auxilliary'' variables which must be added to engineer the interactions \cite{chancellor16a,Leib16a,Dattani18a}. In the case of gate model quantum computers, it is also possible to engineer these higher order interactions out of a sequence of  pairwise interactions, without having to add more qubits. For example, a sequence of CNOT gates can map the total parity of an arbitrary number of qubits to a single qubit value.

In this work, we study the specific case where problems involve higher-than-binary discrete variables, which must be mapped to Boolean variables~\footnote{we use the term Boolean variable here because problem mapping processes, such as minor embedding, parity encoding (i.e.~\cite{Lechner15a}), or error correction codes may mean that the value of a single Boolean variable is stored over multiple physical qubits}. One way to approach this problem is binary encoding, where each value of the discrete variable is assigned a bitstring and the number of binaries used is the minimum allowed by information theory. Another approach is to use a unary encoding for each discrete variable, such that the number of Boolean variables used to encode grows linearly with the size of the discrete variable. The two common methods used here are one-hot and domain-wall encoding \cite{chancellor2019}. 

Both of these encodings are discussed in detail in the next section and a few other examples of encodings are discussed in appendix II. It is worth remarking that while an encoding which only uses unary methods is not a scalable way to build a quantum computer, a setting where the number of (potentially large) discrete variables is increased as the problem size is scaled up does meet the criteria for scalability as discussed in \cite{Blume-Kohout02a}. 

In this paper we consider these encodings at the level of problem mapping, meaning that we assume that the problem should be expressed directly as a QUBO. This is a realistic assumption in the setting of quantum annealing, but may not be for gate model quantum computing, where there may be other compilation methods available. As \cite{Sawaya20a} illustrates, that leads to complicated trade-offs which will not be discussed here. 

The first key result we find is that in the quadratic setting, the domain-wall encoding is the most efficient possible encoding in terms of Boolean variable number per encoded discrete variable (when auxilliary qubits necessary to encode the interaction are accounted for) for practical problems if arbitrary interactions between the discrete variables are desired. This result includes not only one-hot and binary encodings, but also any hypothetical, yet-to-be-discovered encodings.

The second major result of this paper is to gain a better understanding of why the domain-wall encoding experimentally out-performs one-hot in recent quantum annealing experiments \cite{chen21a}. By performing experiments on a D-Wave Advantage 1.1 quantum annealing computer, we find that this is due to a combination of factors, including the fact that fewer Boolean variables induces a smaller solution space, and therefore reduces the effect of thermal fluctuations, as well as the fact that the annealing dynamics effectively ``freeze'' later in the anneal, meaning that the dynamics are more conducive to computation. The second of these effects was theoretically predicted in \cite{chancellor2019}, due to the fact that only a single Boolean variable needs to be flipped to change the value of the discrete variable in the domain-wall encoding, while two need to be flipped in the one-hot setting. These results rely on a simplified version of a problem known as the quadratic assignment problem (QAP), and modeling based on the celebrated Kibble-Zurek mechanism \cite{Kibble76a,Zurek96a}. This mechanism has previously been studied as a model of programmable quantum annealing computers \cite{chancellor16b,Bando20a,Weinberg20a}. In particular, \cite{chancellor16b} used the same model of instantaneous freezing we use here, but for a problem where both the effects of quantum and thermal fluctuations were frozen in. We show that for the experiments here a model based only on thermal fluctuations at the freeze time is a valid approximation.

\section{Background}

\subsection{Flux qubit quantum annealing computer}

\begin{figure}
\begin{centering}
    \includegraphics[width=7 cm]{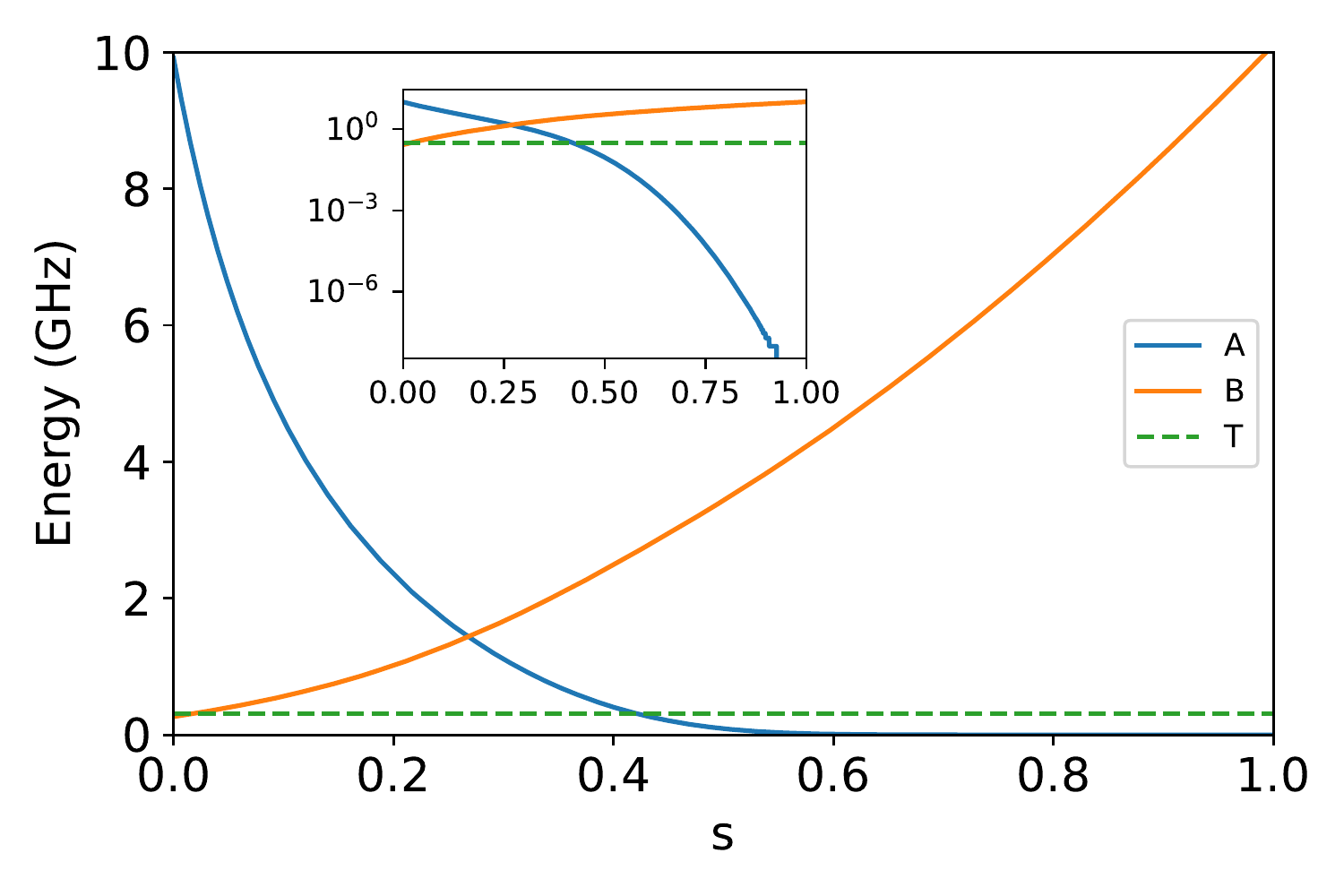}
    \caption{Annealing schedule for the D-Wave Advantage 1.1 processor showing the $A$ (decreasing with increasing $s$), and $B$ (decreasing with increasing $s$) curves along with the temperature ($\approx 15 \, mK \approx 0.31 \, GHz$). The inset is the same plot but with a logarithmic scale on the y-axis.}
    \label{fig:anneal_schedule}
    \par
\end{centering}
\end{figure}

The experiments reported here are performed on a D-Wave Advantage 1.1 programmable flux qubit quantum annealing computer. These devices produce a fully programmable transverse field Ising model on a restricted hardware graph $\chi$. The effective Hamiltonian of the device takes the form 
\begin{equation}
    H(s)=-A(s)\sum_i X_i+ B(s)\left(\sum_{i,j\in \chi}J_{ij}Z_iZ_j+\sum_i h_i Z_i\right) \label{Eq:H_anneal}
\end{equation}
where $X$ and $Z$ are Pauli matrices and $0\le s \le 1$ is the annealing parameter which controls the anneal such that $A(s)$ monotonically decreases and $B(s)$ monotonically increases, as depicted in figure \ref{fig:anneal_schedule}. We only consider forward annealing protocols without schedule variations in this paper, $s=\frac{t}{t_{\mathrm{anneal}}}$ where $t$ is time and $t_{\mathrm{anneal}}=20 \mu s$, which is the default value for these devices. It is worth noting that, while not conventionally included in the Hamiltonian, there is also significant thermal dissipation within the circuit. Since optimal solutions are mapped to low energy states, this dissipation can play a positive role, in fact it was even shown in \cite{dickson13a} that energy increases due to thermal fluctuations can lead to improved performance. There are also negative effects from the interaction with the environment, including thermodynamic effects related to sampling a finite temperature distribution \cite{Albash17a}, the effects of finite temperature play an important role in the results reported here. 

The optimisation problems which these devices solve are programmed into the $B$ terms of equation \ref{Eq:H_anneal}. Equivalently, these problems can be considered as QUBO problems, by realizing that the bit value $b\in \{0,1\}$ can be encoded as $b=\frac{1-Z}{2}$, where the allowed measurement values of $Z$ are $\{1,-1\}$. Note that if we think of the measurement value of $Z$ it can be treated as a classical Boolean variable $\sigma \in \{1,-1\}$, since we are most interested in the higher level picture of the problem not including detailed annealing dynamics we use $\sigma$ for the remainder of the paper. Because the problems we are considering will usually not be subgraphs of the hardware graph $\chi$, we map the problem using minor embedding \cite{choi08a,choi10a}. We use the ``uniform torque compensation'' \cite{uniformTorque} heuristic which is provided as part of the D-Wave Ocean software suite to determine the strength of the minor embedding chains. 

It is important to note that the minor embedding process is likely to have a strong effect on our results, including in some ways which we discuss explicitly. It therefore follows that the results using a problem which did not require minor embedding may be substantially different and probably could not be inferred from our current results. However, we feel that the minor embedded setting is the interesting one for understanding real problems, since it will likely remain an unavoidable step for the forseeable future. While in principle it may be possible to use a set of methods known as parity encoding \cite{Lechner15a,Leib16a,Rocchetto16a} to map problems which are not compatible with the graph, these methods are not commonly used for D-Wave devices, and numerical studies \cite{Albash16b} suggest that minor embedding is likely to be more effective in the D-Wave setting.

\subsection{Unweighted assignment problem \label{sub:uQAP}}

The quadratic assignment problem is an optimisation problem based on a hypothetical situation where $m$ facilities must be assigned to $m$ locations which are each located a set distance $d_{i,j}$ apart, and there is a set flow between each pair of facilities $f_{\alpha,\beta}$. The objective function is the sum of the products of flows and distances $\sum_{i,j} f_{\alpha_i,\beta_j}d_{i,j}$.

For the experimental results presented here, we study an unweighted version of the assignment problem. In other words we consider a version where all combinations of facilities and locations have an equal cost, but each facility must still be assigned a unique location. This version of assignment is no longer a hard combinatorial optimisation problem, since all feasible assignments are equally optimal, and a feasible assignment can be found (for example) by assigning the first facility to a location at random and than iteratively performing assignments based on the lists of remaining facilities and locations. 

While not a direct test of the ability of the annealing computer to solve hard combinatorial optimisation problems, this problem can act as an indirect test, since the ability to find feasible solutions (by which we mean solutions where one facility is assigned to one location and vice-versa) is an important step on the way to finding optimal solutions, such as in optimality certificate-based approaches, e.g., Graver basis \cite{alghassi2019graver}. Moreover, since the unweighted version of this problem is symmetric with respect to permutations of the facilities and locations, it is easier to perform thermal sampling using Monte Carlo methods. This follows from the fact that sampling the solution space around one feasible solution is the same as sampling around any other{--the solution space is a single orbit under the  permutations group action}. 

Moreover, the basic structure of the double constraint, one facility per location, and one location per facility is the same as the underlying constraint in a travelling salesperson (TSP) problem. Consider the placement of the order in which each city is visited in the journey (first, second, third, etc...). Then the constraint that each city can be visited only once is equivalent to saying that each city must have a unique placement in the order of the journey (the same city cannot be both first and third for example). The constraint that each city must be visited at least once can then be enforced by allowing each placement in the order to be used only once (for example only a single city can be third in the order). Since the TSP is not usually thought of as an unweighted assignment with additional quadratic interactions, we discuss this construction explicitly in appendix I.

For the TSP the distances between cities become additional quadratic interactions between the discrete variables. For the (weighted) QAP, quadratic interactions are added between pairs of facility and location choices are added, both cases would add additional couplings which would change the minor embedding of the problem, but since we are primarily interested in the underlying physics of how the device solves the problem, we consider the simple case where only the constraint is mapped, and not additional interactions needed for the problem.

\subsection{Discrete Quadratic Models (DQMs) and encoding to Boolean variables\label{sec:dom_def}}

Many important optimisation problems which exist in the real world involve discrete variables which are higher than binary. In other words, they involve variables for which the allowed values belong to $m$ distinct classes, where $m > 2$. For the QAP, each facility can be placed at any location, so this can also be expressed by discrete variables.  Problems which can be expressed as discrete variables with arbitrary pairwise interactions are conventionally called discrete quadratic models, which we abbreviate as DQM.  

To express the QAP (or TSP) as a DQM, either constraint can be used. For instance, we can either define the DQM variables as the choice of facilities for each location (the order in which each city appears for TSP) or equivalently as the choice of location for each facility (which city appears at each placement in the order for the TSP). In both cases, additional constraints must be added to the DQM to guarantee that the discrete variables take unique values.

Following the convention in \cite{chen21a}, we define DQMs based on a collection of two index variables, $x_{i,\alpha}$, where the index $i$ refers to the variable number, and the index $\alpha$ refers to the variable's class such that

\begin{equation}
    x_{i,\alpha}=\begin{cases} 1 & \text{variable $i$ takes value $\alpha$} \\ 0 & \mathrm{otherwise}. \end{cases}
\end{equation}

An arbitrary DQM is then defined by the following Hamiltonian,
\begin{equation}
    H_{\mathrm{DQM}}=\sum_{i,j}\sum_{\alpha,\beta}D_{(i,j,\alpha,\beta)}x_{i,\alpha}x_{j,\beta} \label{eq:DQM}
\end{equation}
where $D_{(i,j,\alpha,\beta)}$ are the pairwise interactions which determine the overall energy of a configuration. 

An unweighted assignment can be defined by setting $D_{(i,j,\alpha,\alpha)} = 1 \quad \forall\,\alpha,\, i>j$. However it is worth noting that this definition is not unique because $x_{i,\alpha}x_{i,\beta}=0\quad \forall \alpha \neq \beta$, so any finite value of $D_{(i,i,\alpha,\beta)}$ with $\alpha \neq \beta$ does not change the underlying problem. A computationally interesting QAP will also include values of $D_{(i,j,\alpha>\beta,\beta)}=f_{\alpha,\beta}d_{i,j}$ corresponding to each combination of distance and flow and will require $D_{(i,j,\alpha,\alpha)} = \kappa \quad \forall\,\alpha,\, i>j$, where $\kappa$ takes a large enough positive value to enforce the constraint. Similarly a TSP can be defined using the same constraint but taking $D_{(i,j,\alpha>\beta,\beta)}$ values corresponding to distances between cities. By this logic the unweighted assignment problem can also be thought of as an (unphysical) version of the TSP where the distance between all cities is zero (we again refer the reader to appendix I for the details of this mapping).

\paragraph{One-hot constraints:}

Perhaps the simplest way to express a DQM as binary variables is to apply a constraint so that only a single variable can take the value $1$ and enforce that all others take $0$ values. This constraint can be realized as
\begin{equation}
    H_{\mathrm{one\mhyphen hot}}=\kappa \left(\sum_\alpha b_\alpha-1 \right)^2, \label{eq:one-hot}
\end{equation}
where $b_\alpha \in \{0,1\}$ is a binary variable corresponding to each possible value, and $\kappa >0$ is the constraint strength. In this method, if variables also are indexed by $i$ there is a one-to-one correspondence between QUBO variables $b_{i,\alpha}$ and DQM variables $x_{i,\alpha}$. This correspondence makes it tempting to consider the one-hot encoding not as an encoding at all, but simply as constraints on binary variables (this is the way in which we introduced the QAP and TSP in section \ref{sub:uQAP}), each of which correspond to a specific value of a specific variable. However, this way of thinking about problems is no longer useful when no such one-to-one correspondence exists. This is the case for the other two encoding methods we discuss. Since one-hot can also be thought of as an encoding of a DQM (although a rather trivial one), this way of thinking about it is more useful to compare it to other methods on the same footing.  While they will not be directly analyzed here, there are two important extensions to one hot, k-hot constraints and integral encodings, which we review in appendix II.

\paragraph{Domain-wall encoding:}

Another recently proposed method to encode DQMs is the domain-wall encoding \cite{chancellor2019}. In this encoding, for a discrete variable  $m$, only $m-1$ ``spin'' binary variables $\sigma_\alpha \in \{1,-1\}$  (the term ``spin'' derives from condensed matter physics where the direction of the magnetic spin of a spin $\frac{1}{2}$ particle is often denoted in this way). The values encode a domain-wall location (a point where the value of $\sigma$ goes from $-1$ to $1$) in a frustrated segment of a ferromagnetic spin chain. In other words, the Hamiltonian is defined as 
\begin{equation}
    H_{\mathrm{domain \mhyphen wall}}=-\kappa\sum_{\alpha=-1}^{m-2}\sigma_\alpha \sigma_{\alpha+1},\label{eq:domain-wall}
\end{equation}
where the variables are defined for $\alpha \in \{0,m-2\}$ 
and boundary conditions are enforced by setting $\sigma_{-1}=-1$ and $\sigma_{m-1}=1$ ($\kappa$ is again the constraint strength). An advantage of this encoding is that the DQM terms $x_{i,\alpha}$ can be expressed in a way which is linear in the $\sigma$ variables:
\begin{equation}
    x_{i,\alpha}=\frac{1}{2}(\sigma_{i,\alpha}-\sigma_{i,\alpha-1}).
\end{equation}
Because $x_{i,\alpha}$ translates to a linear term in the domain wall encoding, it follows that $x_{i,\alpha}x_{j,\beta}$ will be quadratic in $\sigma$, therefore the domain-wall encoding of a DQM is always also quadratic \emph{in the underlying binary variables}. Note that the notation here is identical to that used in \cite{chen21a} except for the fact that we use $\sigma$ in place of $s$ to avoid confusion with $s=\frac{t}{t_{anneal}}$. It is also worth remarking, that although not the subject of this paper, the domain-wall encoding has potential in simulating quantum field theories \cite{Abel20a,Abel20b}. 

\begin{figure}
\begin{centering}
    \includegraphics[width=7cm]{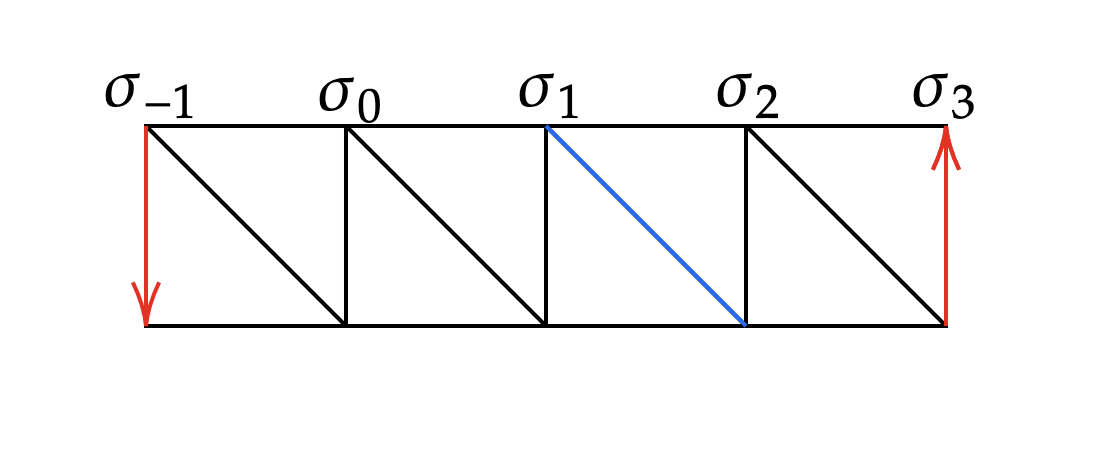}
    \caption{{
    Domain-wall encodings can be represented by triangulations of the Mobius strip. The number of squares represents  the integer number $m$, and the blue twist gives the domain-wall location. Here $m=4$ and the encoded value is $|\downarrow\downarrow\uparrow\rangle$.  
    }
    }
    \label{mobius}
    \par
\end{centering}
\end{figure}

\begin{figure}
\begin{centering}
    \includegraphics[width=7cm]{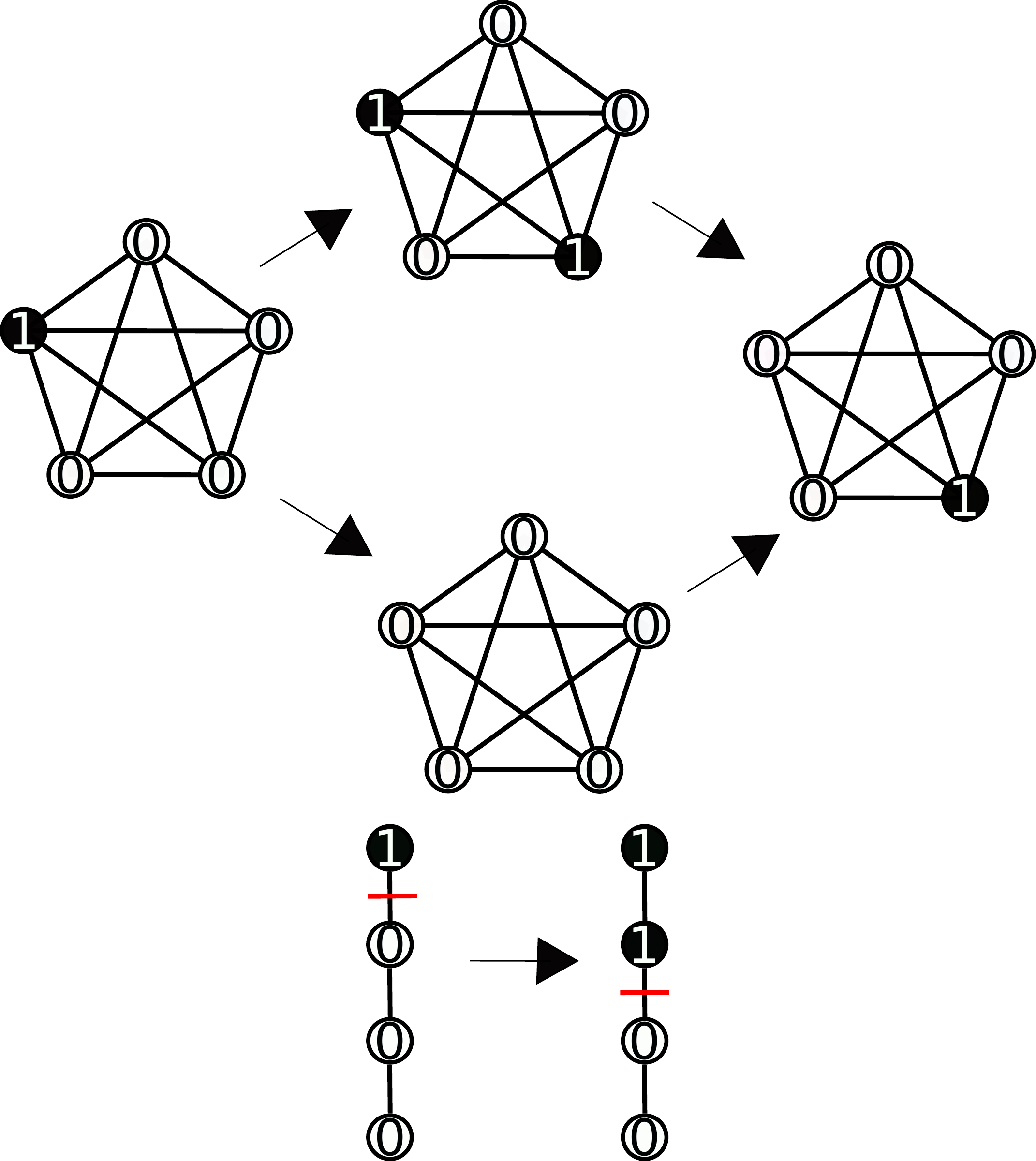}
    \caption{{
    Top: Two ways to update the value of a discrete variable which can take five possible values within a one-hot encoding. Note that there is no way to update the value by changing a single binary variable as all states which are one bit-flip away from a valid state do not represent valid encodings. Bottom: A single bit-flip can update the value of a domain-wall variable. Edges represent interactions between binary variables, and the red line in the bottom subfigure is a guide to the eye to show domain wall location.
    }
    }
    \label{fig:update}
    \par
\end{centering}
\end{figure}

{A pictorial description of the domain-wall encoding is provided by triangulations of the Mobius Strip -- See Figure \ref{mobius}.}  It is also worth observing that unlike one-hot encodings, the value of a discrete encoded by the domain-wall method can be updated by changing a single binary variable. In the case of one-hot these updates require at least two binary variables to be changed as depicted in Figure \ref{fig:update}. This distinction will become important later when we consider the dynamics of physical systems using these encodings, notably it will determine the order of perturbation theory which needs to be considered.

\paragraph{Binary encoding:}

In terms of number of binary variables used to encode a discrete variable (if we ignore auxilliary variables which may be needed to encode interactions), the most efficient method to encode a DQM is to use binary encoding, in other words to assign each value of the discrete variable a bitstring. This method uses $\lceil \log_2 (m) \rceil$ binary variables to encode variables of size $m$. To encode a DQM variable $x_\alpha$, we must assign it a bitstring $r^{(\alpha)}$, where $r^{(\alpha)}_q \in \{0,1\}$ is the value the bitstring takes on binary variable $q$. The encoding of a DQM therefore becomes
\begin{equation}
    x_{i,\alpha}=\prod_{q=0}^{\lceil \log_2 (m) \rceil-1}\left(b_{i,\alpha}r^{(\alpha)}_q+(1-b_{i,\alpha})(1-r^{(\alpha)}_q)\right),
\end{equation}
where again $b_{i,\alpha} \in \{0,1\}$.  

This encoding is not generally quadratic for $m>2$ (it may happen that higher order terms cancel in special cases though). Therefore, in order to translate a binary encoding of a DQM into a QUBO, we must also quadratize higher order terms. In other words, we must transform any terms of the form $b_{i,\alpha}b_{j,\beta}b_{k,\gamma}...$ to one which only has terms of the form $b_{i,\alpha}b_{j,\beta}$. There are multiple methods to translate higher order interactions into quadratic ones, but all require adding at least one auxilliary variable \cite{chancellor16a,Leib16a,Dattani18a}. {The optimal quadratization, i.e., quadratization with the minimum number of auxillary variables, is NP-hard since the problem translates into computing a Groebner basis of a toric ideal \cite{dridi2018novel}}. A fair comparison between binary encoding and other methods such as one-hot and domain-wall, has to include these extra variables. It was shown in \cite{chen21a} that, at least for one approach to quadratization, after these variables are taken into account, binary encoding of the specific colouring problems studied in that paper is less efficient than one-hot or domain-wall. In this paper we show a more general result based on degree-of-freedom counting, that for problems of more than three variables, there is no encoding strategy which can encode general DQM interactions and for which binary encoding uses fewer discrete variables than domain-wall encoding (which in turn uses fewer than one-hot).

\paragraph{Other encodings:}
While we are not aware of any other encoding methods to translate DQMs into QUBOs, it is possible that others could exist which use $n_{\mathrm{var}}$ binary variables where $\lceil \log_2 (m) \rceil<n_{\mathrm{var}}<m-1$. In principle such encodings could always be constructed by taking a binary encoding and adding auxilliary variables which are constrained by the values the variables used in the binary encoding take. For this reason, we consider the possibility of using such encodings in our degree-of-freedom counting analysis as well. Since our analysis is performed from a perspective of counting degrees of freedom, the details of the encoding are not necessary to bound the number of binary variables required.

\subsection{From one-hot to domain-wall}

Since a domain-wall encoding encodes the same information as a set of binary variables which are subject to a one-hot constraint, it is always possible to transform between the two. The recipe is as follows:

\begin{enumerate}
\item Identify all sets of one-hot constrained variables which need to be replaced and translate to a discrete quadratic model (DQM)
\item For each DQM variable, choose an order in which the values will be physically encoded on the chain used for the domain-wall encoding, this is necessary since, while one-hot constraints are permutation symmetric, the domain-wall encoding is not.
\item Generate the necessary domain-wall encodings for each DQM variable.
\item Translate DQM interactions into quadratic and binary terms acting on the new variables.
\end{enumerate}

While it is most intuitive to think of the domain-wall encoding in terms of spin variables $\sigma$ as discussed in section \ref{sec:dom_def}, optimisation problems are usually formulated as QUBOs. To allow easier comparison between the one-hot version (already formulated as a QUBO) and the domain-wall encoding we formulate this encoding as a QUBO as well, using the translation,
\begin{equation}
    \sigma_{i,\alpha}=1-2\,b_{i,\alpha},
\end{equation}
where $\sigma_{i, \alpha} \in \{1,-1\}$ is a spin variable and $b_i \in \{0,1\}$ is a QUBO variable. In this new formulation, we define a DQM variable $x_{i,\alpha}$ as
\begin{equation}
    x_{i,\alpha}=\frac{1}{2}(\sigma_{i,\alpha}-\sigma_{i,\alpha-1})=b_{i,\alpha-1}-b_{i,\alpha}.
\end{equation}
Furthermore the domain-wall constraint can be translated as,
\begin{equation}
    H_{\mathrm{chain}}=-\kappa \left( \sum_{\alpha=-1}^{m-2} 1-2\,b_{i,\alpha}-2\,b_{i,\alpha+1}+4\,b_{i,\alpha}b_{i,\alpha+1}  \right),
\end{equation}
where the pinned boundary values become $b_{i,\alpha=-1}=1$ and $b_{i,\alpha=m-1}=0$. 

The new QUBO generated by this translation has effectively turned half of the one-hot encoding into a domain-wall encoding. The remaining one-hot constraints end up acting like colouring constraints on a fully connected graph, effectively preventing any of the domain-walls from sitting on the same location in the chain and therefore preventing any of the encoded DQM variables from taking the same values.

One might be curious if it is possible to construct an encoding where both sets of one-hot constraints are translated into domain-wall constraints simultaneously, thereby reducing the number of binary variables to $(m-1)^2$ rather than $m(m-1)$. The degree-of-freedom-counting argument in in section \ref{sub:dom_eff} indicates that this is not possible.

\subsection{Freezing of quantum annealing dynamics\label{sub:freezing}}

Before presenting our experimental results it is worth explaining the techniques which we have used to understand the dynamics. In this subsection we mostly provide a qualitative and conceptual explanation, with the details of our techniques being reported in section \ref{sec:methods} In particular, we approximate that the dynamics quickly change from equilibrating very quickly and being well described by a Boltzmann distribution to being completely frozen and experiencing no dynamics. This is fundamentally the approximation which is made behind the celebrated Kibble-Zurek mechanism (KZM) \cite{Kibble76a,Zurek96a,Damski05a,chancellor16b,Bando20a,Weinberg20a} which has been successfullly used to understand a variety of physical systems, both in within the broader Universe and within the laboratory.

Fortunately, the physical temperature of the device is well known, $\approx 15 mK$, so we don't need to measure it. The thermal distribution is not determined by the physical temperature alone, but by the ratio of $\frac{T}{B(s_{\mathrm{freeze}})}$ were $s_{\mathrm{freeze}}$ is the value of the parameter $s$ at which the dynamics freeze. Because the maximum value which we allow $|J|$ to take is $1$, the \textbf{maximum coupling} which is involved in the thermal distribution will have an effective strength of $B(s_{\mathrm{freeze}})$. The quantity $s_{\mathrm{freeze}}$ depends in detail on the dynamics of the system and in general should not be expected to be the same for different encodings. We estimate its value by performing experiments and comparing to simulations and known values as we describe later in this section.

Since the problems we are solving are not in general subgraphs of the connectivity graph of the quantum annealer, we must minor embed them \cite{choi08a}. This introduces the strongest energy scale of the problem since these chains must be strong enough that the interaction terms do not overwhelm the embedding chain couplings which make multiple qubits encode a single binary variable \cite{choi08a}. Since the maximum value of $|J|$ is $1$, we cannot directly increase the strength of these couplers of the minor embedding chains, however, we can weaken the coulplings and fields related to the interactions, which we call the \textbf{QUBO energy scale}. We effectively define the \textbf{chain strength} as the amount by which we scale down these other terms. 

\begin{figure}
\begin{centering}
    \includegraphics[width=7cm]{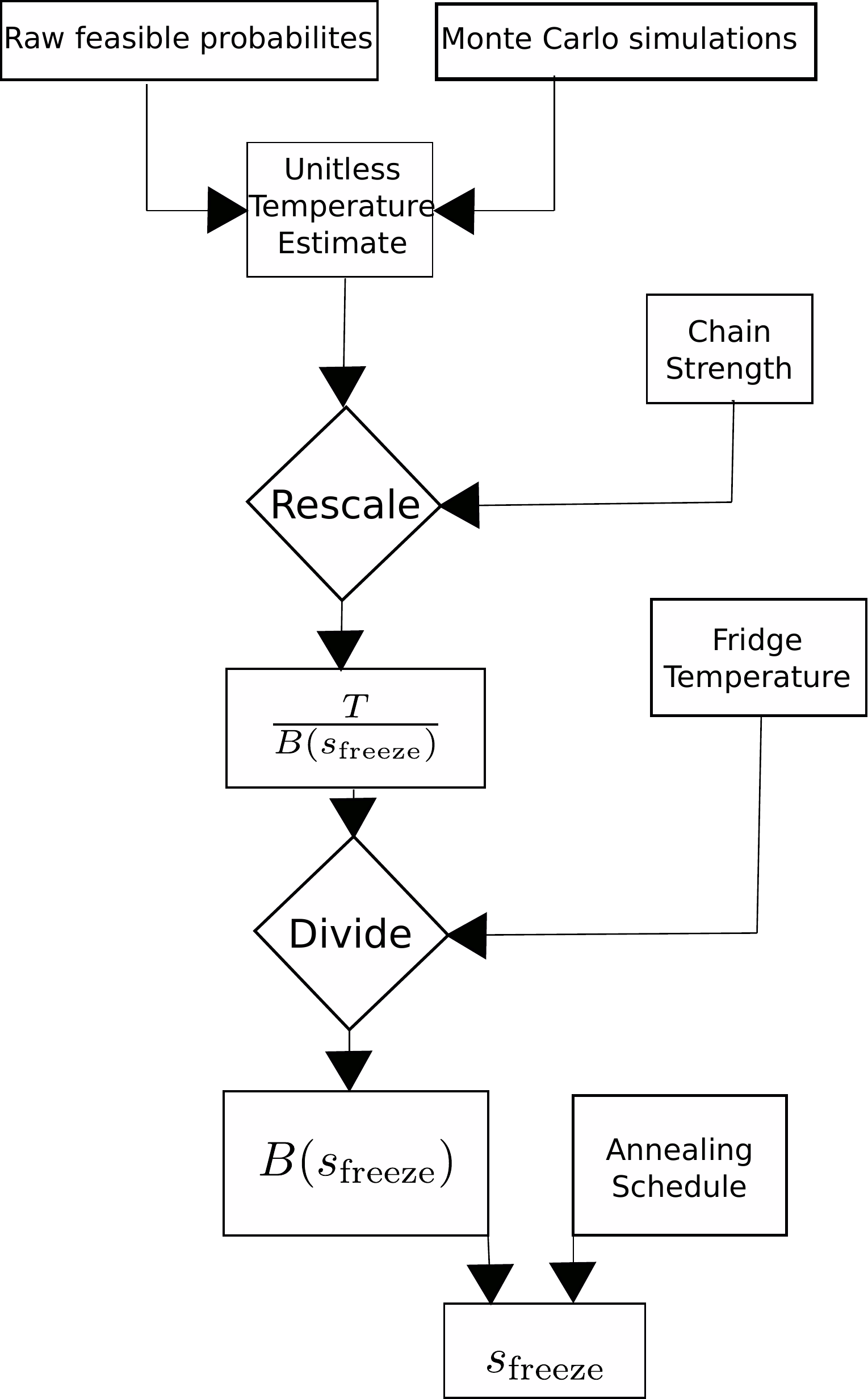}
    \caption{ A flowchart of how we calculate physical quantities related to the freezing of the annealer, specifically included which quantities are incorporated where. This is important to understand the role which quantities such as chain strength and physical device temperature play in the calculation. The initial unitless teperature which is calculated is depicted in figure \ref{fig:T_QUBO}, the values after rescaling with chain strength are shown in figure \ref{fig:T_coup}, and the estimations of $s_{\mathrm{freeze}}$ shown in figure \ref{fig:oh_dom_s},  estimated values of $B(s_{\mathrm{freeze}})$ can be found in figure \ref{fig:oh_dom_E_scales}.
    }
    \label{fig:flowchart}
    \par
\end{centering}
\end{figure}

To estimate $\frac{T}{B(s_{\mathrm{freeze}})}$, which we call the \textbf{effective temperature} and from which can be then used to extrapolate the other quantities, we analyse the \textbf{success probability}, the probability that the annealer found a configuration which satisfies the constraints. By comparing with a numerically estimated thermal distribution (and then rescaling due to minor embedding chain strength, as discussed in the previous paragraph), we are able to extract this quantity since the temperature is known this also gives us the unitful value of  $B(s_{\mathrm{freeze}})$. Since we know the annealing schedule we therefore can also extract $s_{\mathrm{freeze}}$. The process is depicted in the flowchart which appears in figure \ref{fig:flowchart}, the caption of which also references relevant plots in the results section. For this comparison we have chosen to compare the probability of finding a valid solution. While any property should do, based on the KZM picture we are using, this quantity has the advantage of being highly relevant to one of the most important tasks in quantum annealing, finding high-quality solutions to optimisation problems.

\section{Results}

\subsection{Efficiency of domain-wall technique\label{sub:dom_eff}}

While the experimental results reported in this paper all involve the domain-wall and one-hot encodings, it is worth reflecting on whether a direct binary encoding (or some other known encoding) could be more efficient in terms of number of binary variables used, or whether there are room for other, yet-to-be-discovered encodings which could beat the efficiency of the domain-wall encoding. The domain-wall encoding was only proposed in 2019 \cite{chancellor2019}, so it is not unreasonable \textit{a priori} to suspect that other gains could be made. We restrict ourselves to the question of encoding general interactions, in other words encodings which can assign arbitrary energies based on the value of the two variables. This is an important restriction, since binary encoding is known to be efficient for specific interactions, for example in the case of the shortest vector problem \cite{joseph2021}. The structure here which can be exploited is that the interactions can be expressed as $z_iz_j$, where $z_i\in \{0,1...m-1\}$ is the numerical value of each variable, a general interaction would involve all powers of $z_i$ and $z_j$ up to $m-1$, so interactions experessed in terms of these variables would need to be expressed as
\begin{equation}
\sum^{m-1}_{u,v=0}a_{uv}z_i^uz_j^v. \label{eq:interaction_powers}
\end{equation}

We find however for general quadratic interactions that, for problems with four or more variables (as all interesting combinatorial optimisation problems will have), the domain-wall encoding is more efficient in terms number of binary variables required than any known encoding including binary, moreover, we rule out the possibility of any yet-to-be-discovered encoding which is more efficient. We do this using degree-of-freedom counting arguments similar to \cite{chancellor2019}. To start with we consider the case where auxilliary variables are not allowed, in other words, the interaction must be encoded directly in the linear and quadratic degrees of freedom of the binary variables used to encoded discrete variables. We later generalize to the case where auxilliary variables are allowed and find the same conclusion, that the domain wall encoding is the most efficient for all cases of interest. For simplicity we assume that all discrete variables are of size $m$, although in principle the arguments here could be extended to problems containing DQM variables of different sizes. 

An important concept which we use throughout our derivation is the concept of independent degrees of freedom, which we usually just shorten to degrees of freedom. The idea here is that while the effect of QUBO terms on the encoded DQM interaction may not be simple, for every variable within the DQM model every independent element in the tensor from equation \ref{eq:DQM}, there should be at least one QUBO variable (either diagonal or off-diagonal). If there are not sufficiently many such terms, than arbitrary interactions will not be possible.

The reason we have chosen to focus on general interactions is that many interesting real world optimisation problems will not have structure which can be exploited in more specialized encodings, for instance the interactions will generally involve all powers of the variables in equation \ref{eq:interaction_powers}. For example, in a quadratic assignment problem, the quadratic penalties between the variables will not generally have any special relationship which allows any terms of this formula to be eliminated in the interactions, similarly, there is no reason to expect such a relationship between distances between cities in a travelling salesperson problem, or time windows when tasks are performed in a scheduling problem. The only cases we are aware of where a special structure which can be used to make the encoding more efficient is known are when the discrete variables represent numbers (rather than choices) in which case an objective functions which only involve a restricted set of powers of $z$ (in particular the special case where only the first power of  $z$ is involved) could be meaningful and k-hot which is discussed in appendix II and effectively represents a generalisation discrete variables where they can be assigned $k$ values simultaneously. 

There are some cases such as colouring problems, where the interactions are not completely general (in that specific case they prevent coupled variables from taking the same values), but also do not restrict the powers involved when interactions are expressed as equation \ref{eq:interaction_powers}. In these cases, our degree-of-freedom counting arguments do not, strictly speaking, rule out the possibility of a more efficient encoding existing, but to the best of our knowledge, one is not known, so one must revert to general encoding techniques. Given that the more efficient encoding in \cite{joseph2021} and similar works relies on multiplication being easily represented in terms of the binary digits of a number, and there is no analogous structure in terms of a colouring constraint, we find it unlikely that there would be a special structure which would allow colouring problems to be encoded more efficiently, but this cannot be conclusively proven using the methods we present here.

\subsubsection{No auxilliary variables}

In this case all we need to do is to count the linear and quadratic degrees of freedom from the binary variables comprising each discrete variable, and find the size at which there are just enough degrees of freedom to implement the necessary interactions, taking into account that only interactions involving variables used in both terms can induce correlations. We first make an observation, terms which are either linear, or quadratic but between two binary variables which are used to encode the same discrete variable can only contribute a maximum of $m$ degrees of freedom per variable. The reason for this is that the energy shifts these give are not sensitive to the state of the other variable, and therefore can only contribute one degree of freedom per available value. In cases where there are sufficiently many of these degrees of freedom, we then need to show that there are at least $m^2-2m$ quadratic degrees of freedom which contain one binary variable used to encode each.

To understand mathematically why there is a limitation on the number of degrees of freedom which can be contributed by interactions within a single variable encoding, let us consider the general case of two variables of size $m$. In this case the QUBO terms can be written (using two index qubo variables where the first index $i$ or $j$ corresponds to a discrete variable encoding and the second index $\alpha'$ or $\beta'$ corresponds to the qubit number within the discrete variable encoding) as
\begin{widetext}
\begin{align}
    H_{\mathrm{QUBO}}=\sum_{i}^{n_{\mathrm{var}}}\sum_{\alpha',\beta'}B_{(i,j,\alpha',\beta')}b_{i,\alpha'}b_{j,\beta'}\nonumber\\=\sum_{i,j}^{n_{\mathrm{var}}}\sum_{\alpha'}B_{(i,i,\alpha',\beta')}b_{i,\alpha'}b_{i,\beta'}+\sum_{i,j}^{n_{\mathrm{var}}}\sum_{\alpha'<\beta'}(B_{(i,j,\alpha',\beta')}+B_{(i,j,\beta',\alpha')})b_{i,\alpha'}b_{j,\beta'} \label{eq:QUBO_expand} 
\end{align}
\end{widetext}

where $B_{(i,j,\alpha,\beta)}$ is a tensor used to store the QUBO terms used in the encoding in a similar way in which $D_{(i,j,\alpha,\beta)}$ is used to store the interactions between discrete variables and $n_{\mathrm{var}}$ is the number of binary variables used to encode each discrete variable. We now make an observation, if we translate between the QUBO interactions and the energies of the encoded variables, we have
\begin{widetext}
\begin{equation}
    \sum_{\alpha',\beta'}B_{(i,i,\alpha',\beta')}b_{i,\alpha'}b_{i,\beta'} \leftrightarrow \sum_{\alpha<\beta}^{m}D_{(i,i,\alpha,\beta)}x_{i,\alpha}x_{i,\beta}=\sum_{\alpha}^{m}D_{(i,i,\alpha,\alpha)}x_{i,\alpha}
\end{equation}
\end{widetext}
the second equality can be derived from the fact that each variable can only take one value, so therefore $x_{i,\alpha}x_{i,\beta}=0$ for all cases where $\alpha \neq \beta$, along with the fact that $(x_{k,\alpha})^2=x_{k,\alpha}$. The result is that, no matter how many $B_{(i,j,\alpha,\alpha)}$ degrees of freedom are available, they can only contribute a maximum of $m$ independent degrees of freedom for each discrete variable, since there are only $m$ possible values of $x_{k,\alpha}$. Equation \ref{eq:dof_breakdown} shows where the different degrees of freedom come from for an interaction between two variables, giving a total of $2\min(m,\frac{n_\mathrm{var}(n_\mathrm{var}+1)}{2})+n_{\mathrm{var}}^2$ degrees of freedom for an interaction between two discrete variables of size $m$ indexed with $\alpha$ and $\beta$ each encoded into $n_{\mathrm{var}}$ binary variables.
\begin{widetext}
\begin{equation}
    H_\mathrm{QUBO}=\underbrace{\sum_{ \alpha',\beta'}\sum_{i}^{n_{\mathrm{var}}}B_{(i,i,\alpha,\beta')}b_{i,\alpha'}b_{i,\beta'}}_{\scalebox{1.25}{$2\min(m,\frac{n_\mathrm{var}(n_\mathrm{var}+1)}{2})$}}+\underbrace{\sum_{i,j}^{n_{\mathrm{var}}}\sum_{\alpha'<\beta'}(B_{(i,j,\alpha',\beta')}+B_{(i,j,\beta',\alpha')})b_{i,\alpha'}b_{j,\beta'}}_{\scalebox{1.25}{$\quad \quad  \quad n_{\mathrm{var}}^2$}}. \label{eq:dof_breakdown}
\end{equation}
\end{widetext}

Put another way, we can divide the terms into two categories. The first we call \textbf{correlating} and are represented by the second set of terms in equation \ref{eq:dof_breakdown}, these terms necessarily involve binary variables used to construct each of the two DQM variables, as such these terms can create energy correlations between the two DQM variables in principle there can be up to $m^2$ independent correlating terms. On the other hand, we call the first set of terms \textbf{non-correlating}, since they involve binary variables which are used to construct only one of the two DQM variables. Non-correlating terms cannot induce energy correlations between the DQM variables by construction. While non-correlating terms can in general be constructed from correlating terms, the converse is not true, there is no way to represent a correlated energy using a sum of terms which only each contribute to the energy of one of the two DQM variables. When counting the non-correlating terms we have to be careful not to over-count redundant terms which could be constructed from the others. The interesting quantity is therefore how many \emph{independent} (in the sense that the do not represent a contribution to the DQM energy which can be constructed from terms which are already present) degrees of freedom these terms contribute. If we examine each DQM variable independently, we find that each will have $m$ possible energies for the $m$ possible values the variable can take, and therefore only $m$ indpendent terms can be constructed. Since these are the only energies to which non-correlating terms can contribute, than no matter how many non-correlating terms are available, only $2m$ of them will ever be independent of each other.

As an example of the limitation on the number of degrees of freedom which the non-correlating terms can contribute, consider the particular case of domain-wall encoding. Terms of the form $b_{i,\alpha}b_{i,\alpha}=b_{i,\alpha}$ will uniquely define the energy of the $m$ values of a DQM variable up to an irrelevant offset, while terms of the form $b_{i,\alpha}b_{i,\beta}$ with $\alpha \neq \beta$ represent interactions between different variables on the same Ising chain, these terms will contribute an energy only if both variables are $1$ as opposed to $0$, and therefore if $\alpha>\beta$ would be exactly equivalent to $b_{i,\alpha}$, and therefore not contribute an independent degree of freedom.

We start our anlysis by counting degrees of freedom within the domain-wall encoding, which uses $n_{\mathrm{var}}=m-1$ binary variable to encode a discrete variable of size $m$. Firstly, for $m>2$, we find that indeed $\frac{(m-1)(m-2)}{2}>m$, so there are sufficient terms to gain $m$ degrees of freedom per variable. Next we observer that $(m-1)^2=m^2-2m+1>m^2-2m$, so again sufficient degrees of freedom, but with only one spare. 

Next, we ask what happens when we remove one binary variable from even just one of the discrete variable encodings, we now have $(m-1)(m-2)=m^2-3m+2$ degrees of freedom, for $m>2$ we find that there are insufficiently many degrees of freedom (and will be even fewer if the number of binary variables are reduced further). We therefore conclude that unless auxilliary variables are added for each coupling, there is no way to construct a more efficient general interaction than the domain-wall encoding. Note that this result is somewhat more substantial than the one in \cite{chancellor2019}, which showed that the degrees of freedom \emph{which were used in the encoding} were just sufficient, here we have shown that the domain-wall encoding is maximally efficient even if all linear and quadratic interactions are used. In this section we only consider the case of arbitrary quadratic interactions between discrete variables, while the arguments here could in principle be extended to higher order interactions (those involving more than $2$ discrete variables) this is beyond the scope of the present work.

\subsection{Including auxilliary variables}

We now ask whether more efficient encodings are possible if we allow each interaction to be supplemented by additional auxilliary variables which could, for example, be used to engineer higher-than-quadratic interactions, as shown in \cite{chancellor16a,Leib16a,Dattani18a}. Unlike the previous case, where the number of Boolean variables scales only with the number of discrete variables, in this context the auxilliary variables will be required for each interaction, therefore the degree of the interaction graph for the discrete variables becomes important. Our strategy proceeds as follows:
\begin{enumerate}
\item Count the number of degrees of freedom which are deficient in each interaction, and therefore how many need to be added with auxiliary variables
\item Derive an expression for how many degrees of freedom will be contributed by adding $n_{\mathrm{aux}}$ auxilliary variables in encoding an interaction
\item From this expression, derive an expression for the critical interacting graph degree $d_{\mathrm{crit}}$ above which a method using auxilliary variables cannot be more efficient than domain-wall even if  all degrees of freedom can be used optimally in terms of the number of variables and the size $m$ of the discrete variables (assuming all are the same size)
\item Evaluate this expression numerically for sizes up to $m=1000$ for various number of Boolean variables per discrete variable ($n_{\mathrm{var}}$) in the allowed range (binary encoding up to one fewer than the number required for domain-wall)  
\item Based on these results, argue that for all cases of practical interest, specifically those with the number of discrete variables $n_{\mathrm{var}}>3$, there cannot be encodings which make use of auxilliary variables and are more efficient than domain wall
\end{enumerate}

Since there are many different potential strategies for using auxilliary variables, and we do not want to consider every possibility independently, we instead base our argument on counting degrees of freedom, if there are insufficient degrees of freedom to independently control the energy of every configuration of a pair of variables, than \emph{no} strategy can be used to construct general interactions\footnote{We are not sure if the converse of this statement is true, in other words, we aren't aware of a construction which always works when sufficient degrees of freedom are available. Calculations done in this way should therefore be viewed as bounds}. We first use the results of the previous section to count the degrees of freedom before any auxilliary variables are added, if each discrete variable of size $m$ uses $\lceil \log_2(m) \rceil \le n_{\mathrm{var}}<m-1$ binary variables than there will be 
\begin{equation}
    D=m^2-n^2_{\mathrm{var}}-2\min(m,\frac{n_{\mathrm{var}}(n_{\mathrm{var}}+1)}{2}) \label{eq:missing_dof}
\end{equation}
degrees of freedom which will need to be added via linear terms on auxilliary variables and interactions between auxilliary variables. We will now assume that for each interaction between a pair of discrete variables $n_{\mathrm{aux}}$ auxilliary variables are included, which interact both with the discrete variables and between themselves. The goal of adding these variables is that each of them can be constrained to represent additional derived properties of the binary variables used to encoded discrete variables, such as ``majority votes'' or parity values of subsets. Since the interactions between the auxilliary variables and the discrete variables will be used to control the values the auxilliaries take, they should not be counted toward the total available degrees of freedom, but both linear terms on the auxilliaries, and quadratic terms between them should, meaning that $n_{\mathrm{aux}}$ auxilliary variables give us $\frac{n_{\mathrm{aux}(n_\mathrm{aux}+1)}}{2}$ additional degrees of freedom. 

We can therefore calculate the minimum necessary number of auxilliary variables needed to make up for the missing degrees of freedom by setting
\[
    \frac{n_{\mathrm{aux}}(n_{\mathrm{aux}}+1)}{2}\ge D. 
\]
By completing the square, we find the minimum number of auxilliary variables required to generate enough degrees of freedom
\begin{equation}
    n_{\mathrm{aux}}=\left\lceil \sqrt{2}\sqrt{D+\frac{1}{8}}-\frac{1}{2} \right\rceil,\label{eq:aux_num}
\end{equation}
where the bracketing symbols indicate ceiling, since this value must be an interger. The average number of binary variables used per discrete variable than becomes 
\begin{equation}
n_{\mathrm{bin}}=n_{\mathrm{var}}+\frac{d}{2}n_{\mathrm{aux}}, \label{eq:n_bin}
\end{equation}
where $d$ is the average degree of the interaction graph of the problem. The condition for an encoding to be more efficient than the domain-wall encoding is therefore $n_{\mathrm{bin}}<m-1$, rearranging, we can find a critical degree below which an auxilliary variable based encoding could be more efficient than the domain-wall encoding, this yields
\begin{equation}
    d_{\mathrm{crit}}=2\frac{m-1-n_{\mathrm{var}}}{n_{\mathrm{aux}}}. 
\end{equation}
Substituting in variables yields
\begin{equation}
    d_{\mathrm{crit}}=\sqrt{2}\frac{m-1-n_{\mathrm{var}}}{\sqrt{m^2-n^2_{\mathrm{var}}-2\min(m,\frac{n_{\mathrm{var}}(n_{\mathrm{var}}+1)}{2})+\frac{1}{8}}-\frac{1}{2}}. \label{eq:d_crit}
\end{equation}
while applying this formula by hand to check every allowed value of $n_{\mathrm{var}}$ for a given value of $m$ is impractical, especially for larger $m$, this value can be readily calculated by a computer. It is furthermore clear that in the limit of large $m$ and small $n_{\mathrm{var}}$, in other words for binary encodings with large $m$, the limiting value is $d_{\mathrm{crit}}=\sqrt{2}$. 

\begin{figure}
\begin{centering}
    \includegraphics[width=7 cm]{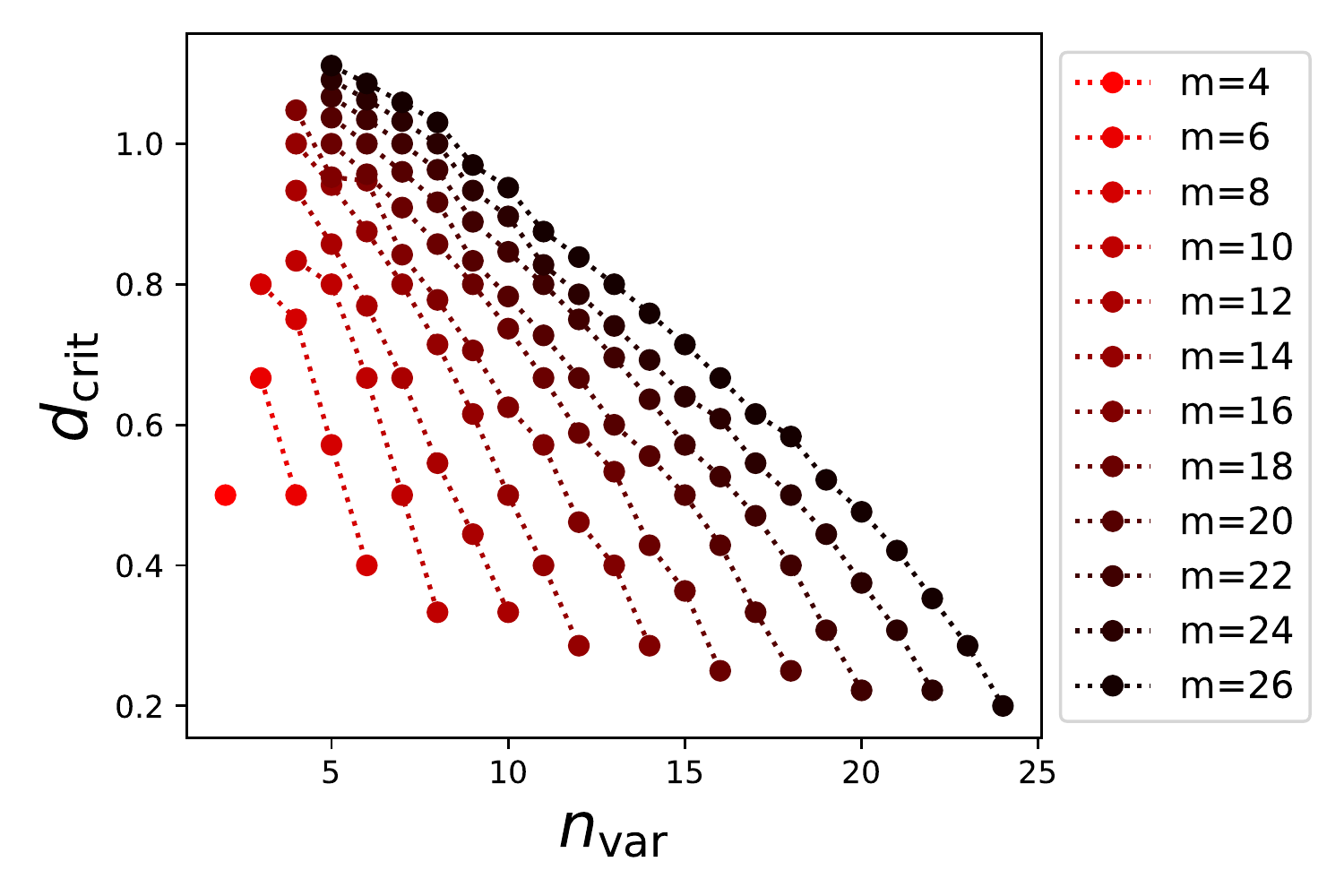}
    \caption{Critical degree $d_{\mathrm{crit}}$ as defined in Eq.~\ref{eq:d_crit} for encodings using auxilliary variables with $n_{\mathrm{var}}$ binary variable per discrete variable for discrete variables sizes in the range $m=4$ to $m=26$. Only even values are plotted for easier visual interpretation, no significant even-odd effects have been observed. At all sizes shown here, choosing the minimum value of $n_{\mathrm{var}}$ (corresponding to a binary encoding) yields the highest $d_{\mathrm{crit}}$ and therefore the most efficient encoding, although not shown, a similar trend has been observed up to $m=1000$.}
    \label{fig:dcrit_nvar}
    \par
\end{centering}
\end{figure}

\begin{figure}
\begin{centering}
    \includegraphics[width=7 cm]{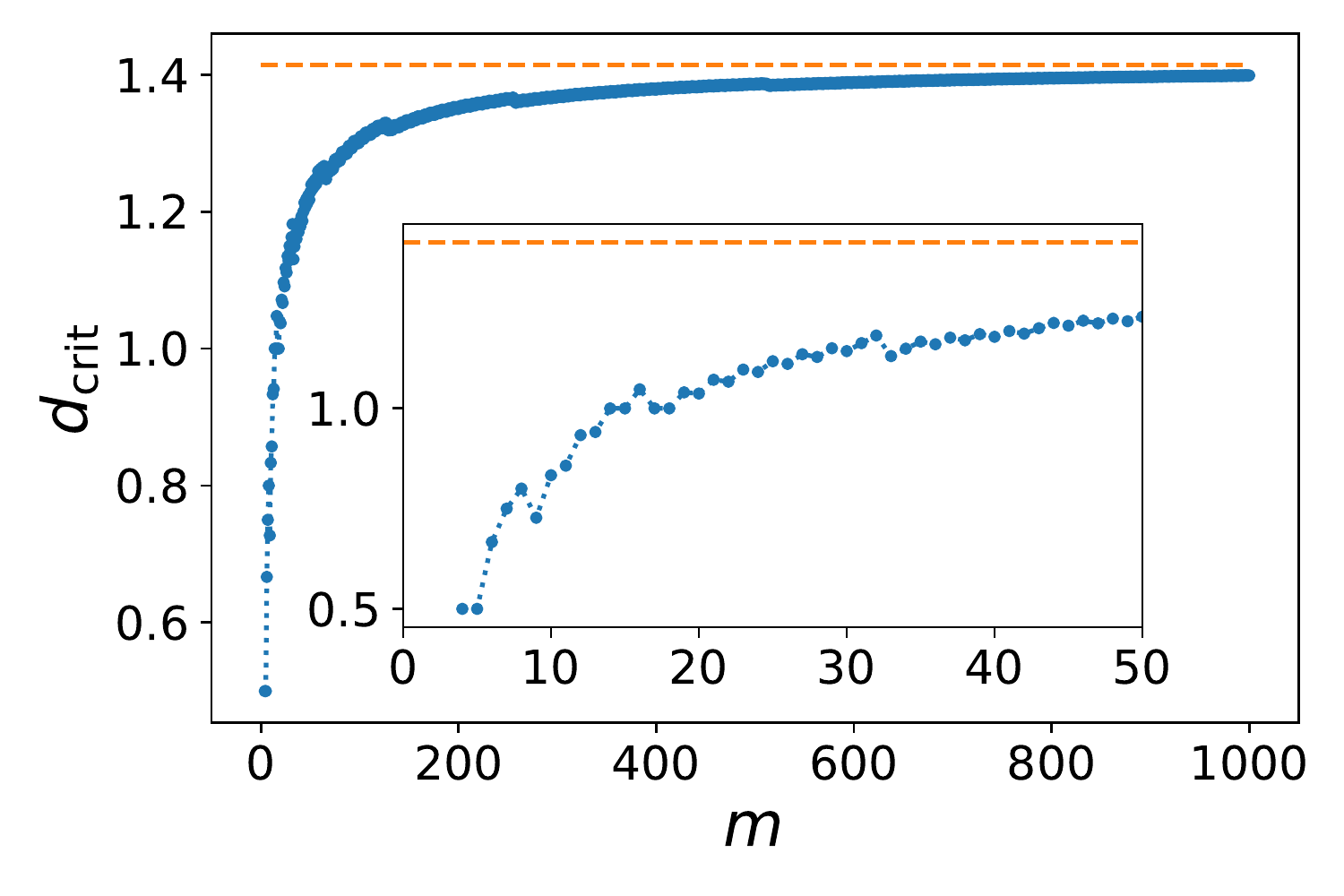}
    \caption{Critical degree $d_{\mathrm{crit}}$ as defined in Eq.~\ref{eq:d_crit} for optimal encoding using auxilliary variables versus the size of the variable. Dashed line is a guide to the eye located at $d_{\mathrm{crit}}=\sqrt{2}$. The inset is a zoom of the outer plot. All values of $\lceil \log_2(m) \rceil \le n_{\mathrm{var}}<m-1$ were tested and the highest value of $d_{\mathrm{crit}}$ attained is plotted here, we observe that this always corresponded to the binary encoding case, $n_{\mathrm{var}}=\lceil \log_2(m) \rceil$.}
    \label{fig:dcrit_m}
    \par
\end{centering}
\end{figure}

From applying eq.~\ref{eq:d_crit} for all values $3<m\le 1000$ we find firstly that the highest values of $d_\mathrm{crit}$ are always attained for binary encodings, in other words where the minimum number of binary variables are used to encode each discrete variable, this trend is shown for small $m$ in fig.~\ref{fig:dcrit_nvar}. As fig.~\ref{fig:dcrit_m} shows the maximum $d_\mathrm{crit}$ asymptotically approaches $\sqrt{2}$ from below. 

Let us now consider what this critical value of degree means for real optimisation problems. Firstly, we are only interested in the connected components of the interaction graphs, since disconnected components can be solved separately. This limits the minimum average degree a graph can have based on the number of nodes (each of which correspond to DQM variables in this case), since the minimum connected graph with $q$ nodes is either a star or line graph with $q-1$ edges, and therefore an average degree of $d_q=2\frac{q-1}{q}$. For example, a three node connected graph cannot have a degree less than $\frac{4}{3}$ and a four node connected graph cannot have degree less than $\frac{3}{2}$, since $\frac{3}{2}>\sqrt{2}> d_\mathrm{crit}$, it follows that an auxiliary based encoding cannot be more efficient than the domain-wall encoding for problems on graphs with more than three nodes (and therefore containing more than three DQM variables). As discussed in \cite{Blume-Kohout02a} increasing the variable size with a fixed number of variables is not a scalable way to perform quantum computation, so therefore the domain-wall encoding is the most efficient encoding in terms of binary variable count for problems needing general interactions in all cases of practical interest. 

 It is worth remarking that while this result only applies to arbitary interactions between the discrete variables, the interactions between the variables are allowed to take essentially arbitrary values in both the QAP and the travelling salesperson problem, so our results do imply that the domain-wall encoding is the most efficient way to encode both of these problems. 

\subsection{Domain wall versus one-hot for quantum annealing}

\begin{figure}
\begin{centering}
    \includegraphics[width=7 cm]{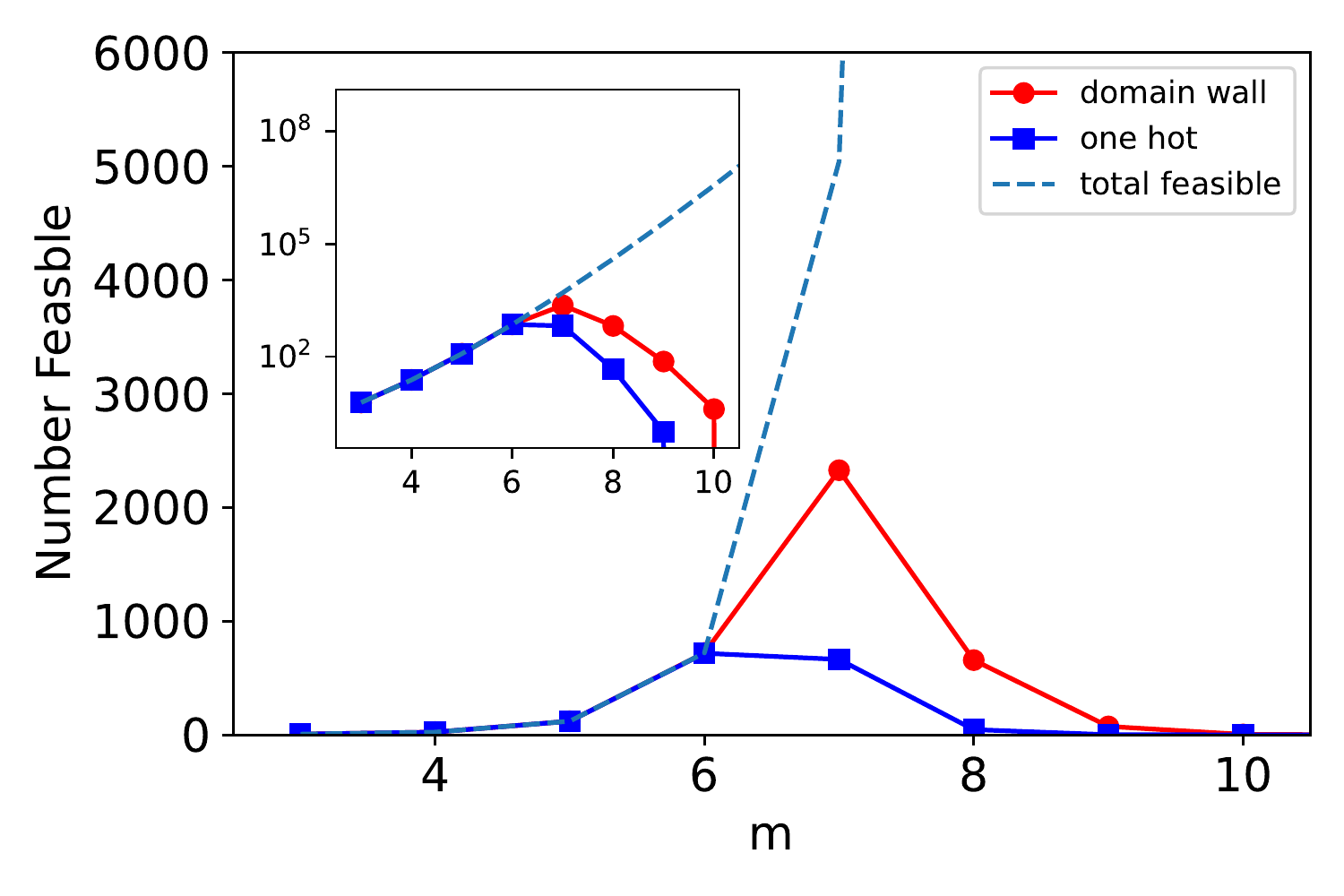}
    \caption{Number of unique feasible solutions found for both encodings over all $10$ embeddings versus $m$. Dashed line shows the total number of feasible solutions at each $m$ value. The inset is the same plot but with a logarithmic scale on the y-axis. Note that no matter how effective a solver is it can never find more than the total number of feasible solutions (indicated by the dashed line). The relevant performance measure is therefore how close the device gets to the best possible performance.
    }
    \label{fig:num_feas_expt}
    \par
\end{centering}
\end{figure}

\begin{figure}
\begin{centering}
    \includegraphics[width=7 cm]{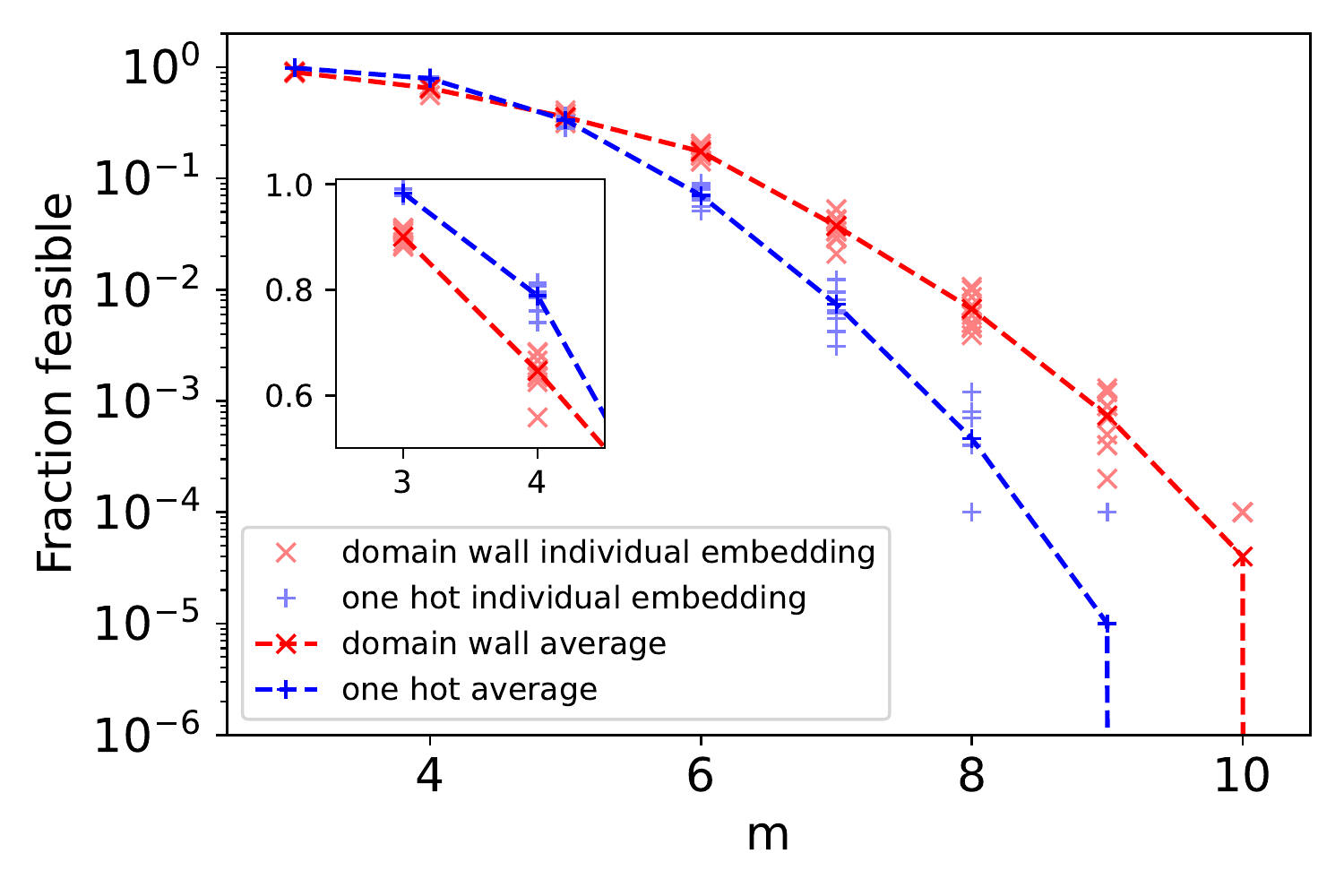}
    \caption{Fraction of returned solutions which are feasible at different values of $m$. Full colour symbols indicate average over all $10$ embeddings, while fainter symbols indicate values for individual embeddings. The inset shows a zoom on part of the plot with a linear y-axis to show the behavior before the crossover more clearly. }
    \label{fig:frac_feas_expt}
    \par
\end{centering}
\end{figure}

Now that we have established the domain-wall encoding as the optimal encoding for general DQM's we turn to the other main result of this paper, asking whether the annealing dynamics are favourable for a domain-wall encoded problem when compared to the same problem with a one-hot encoding. To do this,  we first run the one-hot and domain-wall encoding of the unweighted assignment for $10$ different embeddings with $10,000$ anneals for each.  As figure \ref{fig:num_feas_expt} shows, the domain-wall encoding is always able to find the same number or more feasible solutions in our experiments. However, as figure \ref{fig:frac_feas_expt} shows, there is a crossover in terms of the probability that an individual annealing run yields a feasible solution, with one-hot performing better at smaller sizes and domain-wall performing better as the problem becomes larger. This is interesting because, while not in a particularly interesting regime (small sizes), this is the first case we are aware of where one-hot is able to outperform domain-wall by any metric. As we discuss later the cause of this crossover can be explained by a simple thermodynamic model. It is worth remembering that the purpose of this analysis is not to directly test the ability of the device to solve hard combinatorial optimisation (unweighted assignment is not a computationally hard problem), but rather to perform analysis on the underlying physical dynamics which allow the device to find solutions which satisfy a constraint.

In section \ref{sub:eff_temp} in the spirit of the Kibble-Zurek Mechanism, we use these finite temperature models to estimate an effective temperature, and therefore determine which encoding yields dynamics which are more favourable to computation. We find that both thermodynamic and dynamical effects contribute to the superior performance of the domain-wall encoding. We also find that the chain strength chosen by a standard heuristic is actually lower for the one-hot encoding. 

\begin{figure}
\begin{centering}
    \includegraphics[width=7 cm]{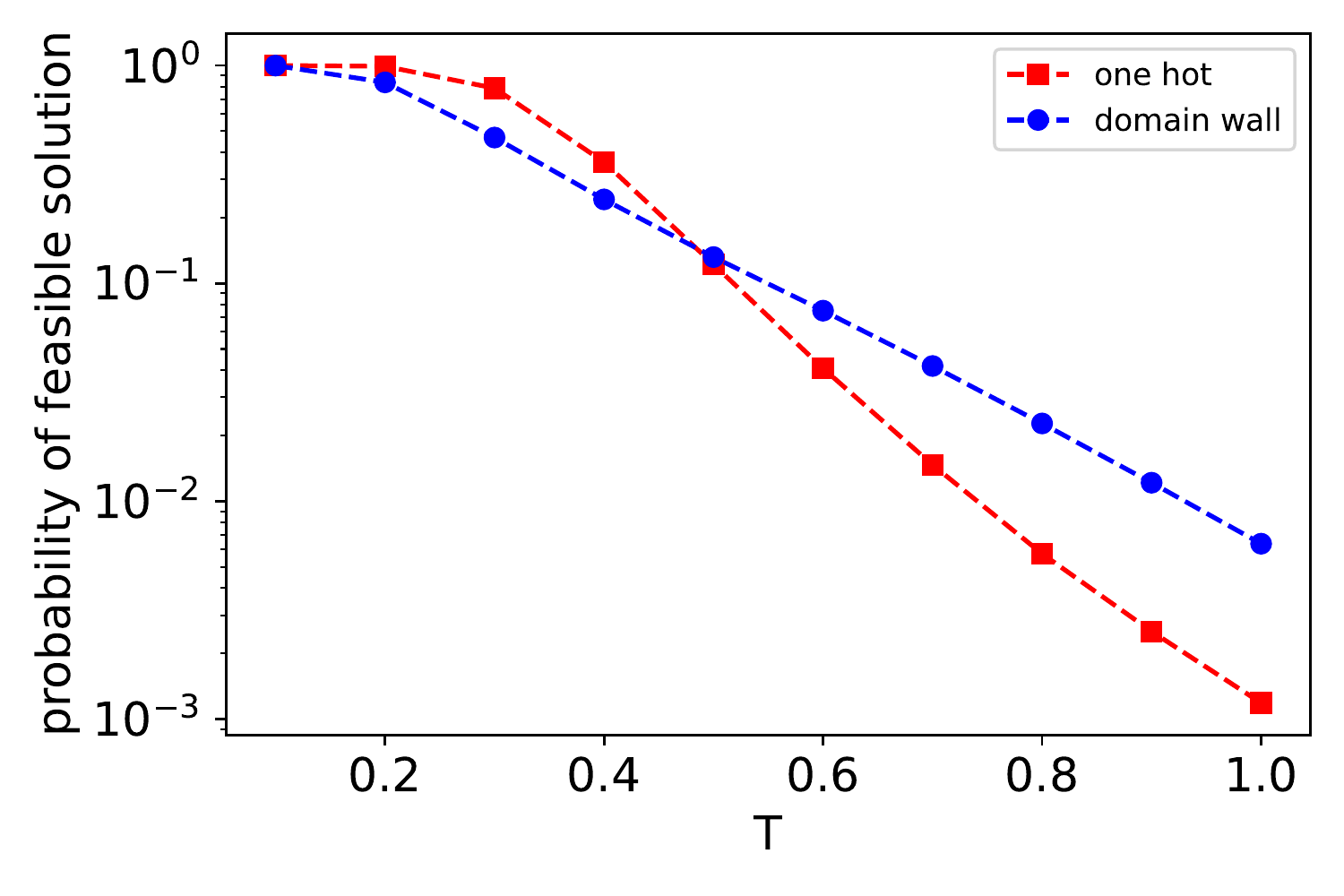}
    \caption{Comparison of thermal probabilities for feasible solutions  at different temperatures for domain-wall and one-hot encodings of $m=8$ unweighted assignment problem. Temperature on this plot is in dimensionless units. All points represent Monte Carlo sampling with $10^7$ samples, standard error and 95\% error bars are both much smaller than the depicted symbols.
    }
    \label{fig:thermal_oh_dw_m8}
    \par
\end{centering}
\end{figure}

To understand the thermodynamics of the problems better, we first examine how temperature effects the QAP when using the domain-wall or one-hot encodings. To do this we perform a simple Metropolis algorithm on the two encodings at various temperatures for $m=8$. While it won't necessarily be true for every problem, these converge well for the unweighted assignment problem. (Convergence of the Monte Carlo sampling is discussed in appendix III.) As Fig.~\ref{fig:thermal_oh_dw_m8} shows, at lower temperature the one-hot encoding performs better, but the domain-wall encoding performs better above a unitless temperature of $T\approx 0.5$. It is expected that at high temperatures, the domain-wall encoding should perform better, since the encoding uses $m$ fewer qubits, and therefore the solution space is $2^m$ times smaller, $256$ times in the $m=8$ case examined here

\begin{figure}
\begin{centering}
    \includegraphics[width=7 cm]{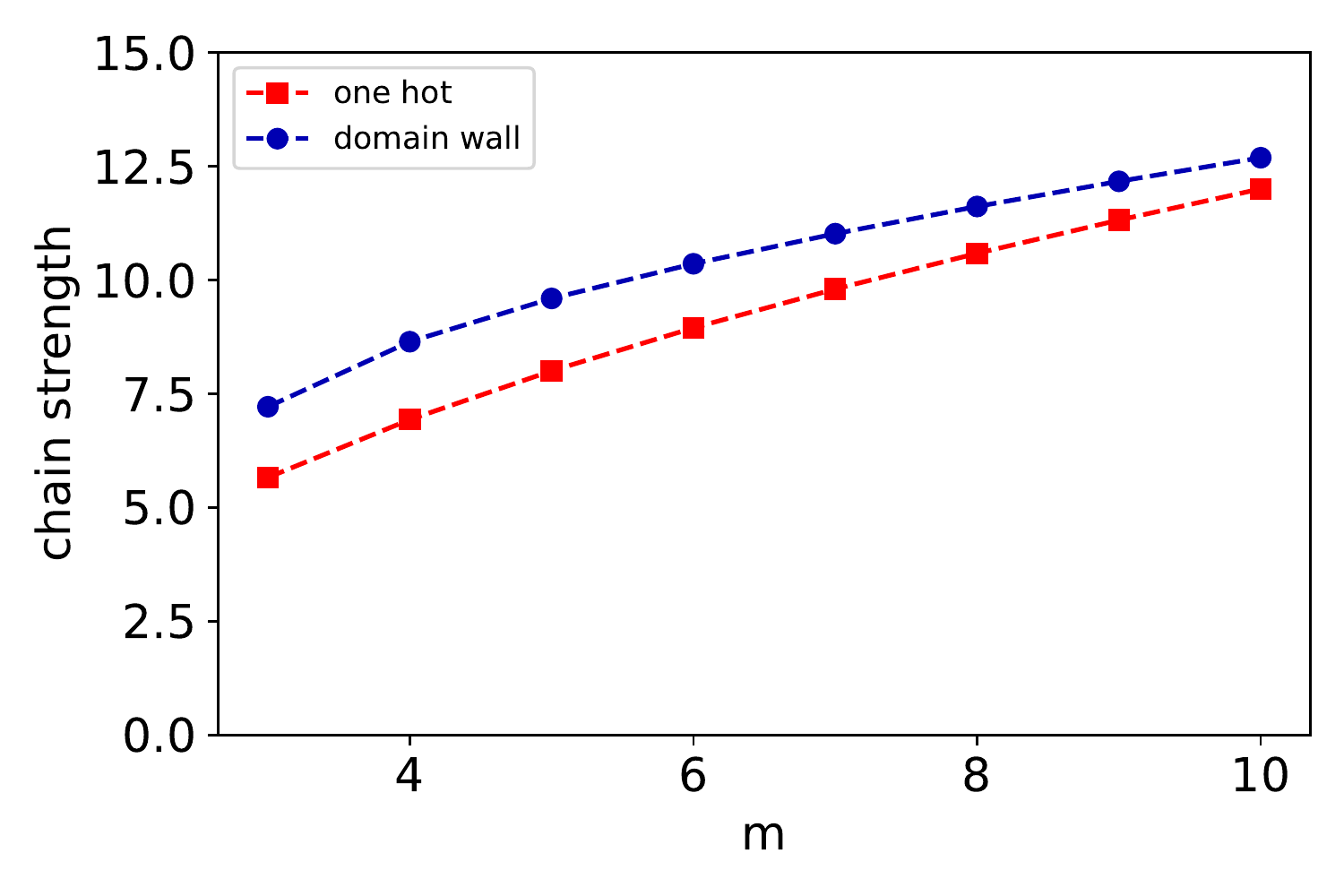}
    \caption{Embedding chain strength calculated using the default uniform torque compensation heuristic for domain-wall and one-hot encodings. Because of dynamaic range limitations, the inverse of this quantity effectively sets the value of $\kappa$ in equations \ref{eq:one-hot} and \ref{eq:domain-wall}.}
    \label{fig:uniform_torque_m_dw_oh}
    \par
\end{centering}
\end{figure}

However, even assuming that the device always samples an equilibrium distribution at the same temperate, there are other considerations for device performance. To start with, the problem must be minor embedded in both cases, and the embedding chain strength required may not be the same for both encodings. Although experimentally we have seen that chain breaks are rare (this is also predicted by the large relative energy scale for the chains seen in figure \ref{fig:oh_dom_E_scales}), dynamic range limitations mean that the stronger the embedding chains needed, the higher the effective temperature of the encoded problem. For this study we have used the ``uniform torque compensation'' heuristic available in the Ocean software suite with default parameters \cite{uniformTorque}. Note that this heuristic does not depend on the embedding itself, only the pre-embedding problem. This heuristic estimates the necessary chain strength based on  summing the couplings each logical variable experiences in quadrature. As fig.~\ref{fig:uniform_torque_m_dw_oh} shows, the heuristic consistently assigns stronger chain strengths for the domain-wall encoding, although the difference decreases as $m$ increases. While beyond the scope of this investigation, it may be fruitful to examine whether different heuristics are more appropriate when domain-wall encoding is used, since the current heuristic was likely tested and tuned based on problem structures which are currently used, likely including one-hot constraints. 

\begin{figure}
\begin{centering}
    \includegraphics[width=7 cm]{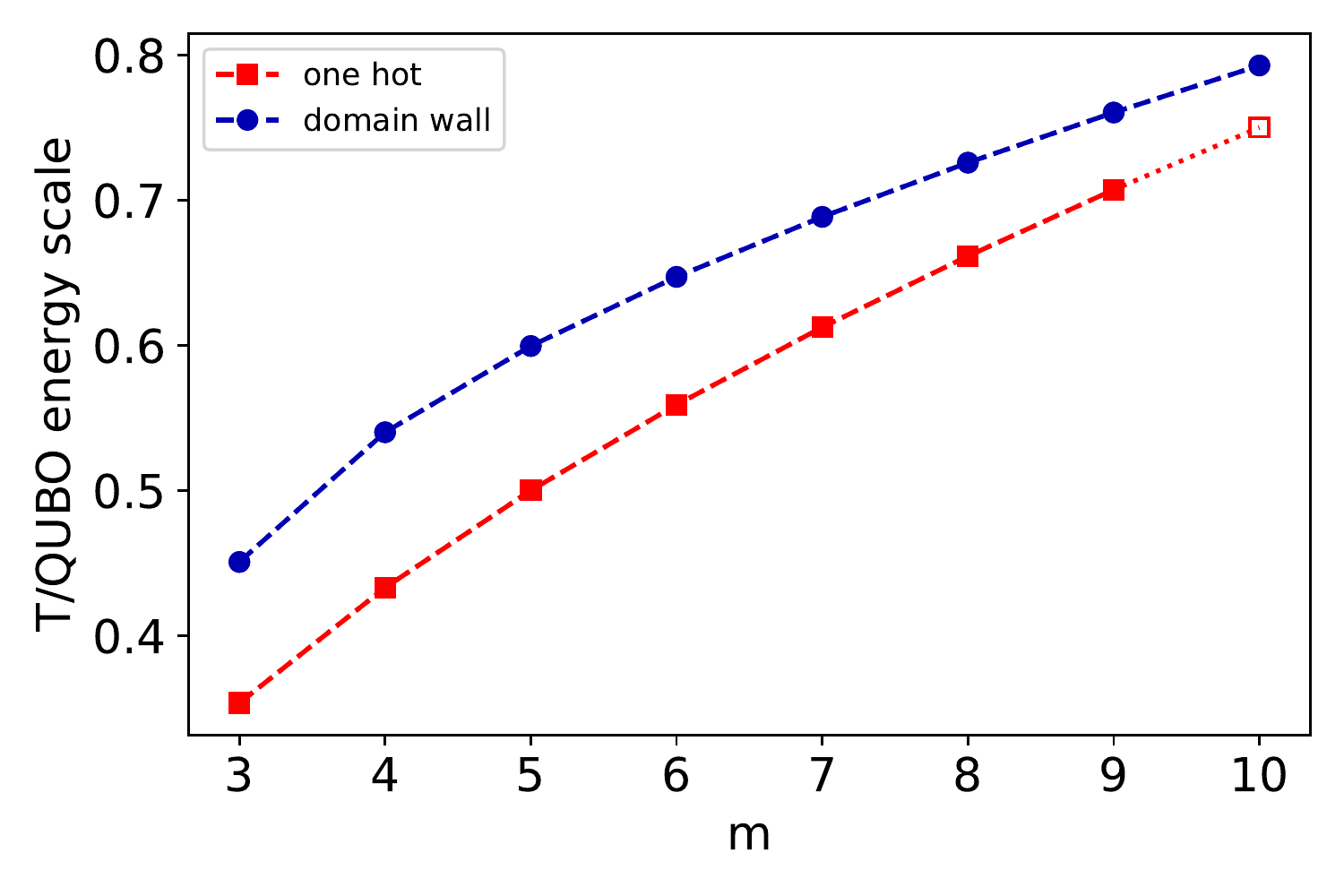}
    \caption{Estimated effective temperature for one-hot and domain-wall encodings at different values of $m$. These values are calculated using a rough estimate of the unitless temperature and rescaling by the minor embedding chain strength. Since the heuristic has chosen a higher chain strength for the domain-wall encoding, this results in a higher effective temperature (unitless temperature of $0.063$). 
    }
    \label{fig:effective_temp_embed}
    \par
\end{centering}
\end{figure}

The final ingredient to investigate equilibrium effects on the annealing computer is an estimate of the temperature. While the physical temperature of these devices is well known at around $15 \,mK$, the energy scale of the couplers at the point where the dynamics freeze is harder to estimate. This energy scale plays an important role, because the relative probability of configurations in a Boltzmann distribution is not solely determined by the physical (unitful) temperature but by a unitless ratio of the temperature of the energy scale (temperature already in units of $mK$, so Boltzmann constant is not necessary). From equation \ref{Eq:H_anneal} we find that $B(s_{\mathrm{freeze}})$ term determines the energy scale of the couplers, where $s_{\mathrm{freeze}}$ is the value of $s$ at the freezing point. In general we expect $s_{\mathrm{freeze}}$ to be problem dependent. For our final anysis we want to estimate it experimentally, but before we do, it is instructive to see how the system behaves if we make a very rough approximation that for both encodings $B(s_{\mathrm{freeze}}) \approx 5 \,GHz$.

While this value is very approximate, and in practice the energy scale will change with the freeze time, choosing an approximate value of the correct order of magnitude for the device will allow investigation of relevant qualitative effects. Using this energy and physical temperature to calculate the the overall unitless temperature, we find $T=0.063$. 
In section \ref{sub:eff_temp} we use our modeling techniques to estimate the effective unitless temperature of the different encodings. However, at this stage an approximate value is sufficient to demonstrate the effects we want to show. Since some of the dynamic range is taken up by the embedding chains \cite{Albash17a} 
 the effective temperature we should use to model the solver needs to be further rescaled to take this into account, by about a factor of $10$ in the $m=8$ case. Fig.~\ref{fig:effective_temp_embed} shows the effective unitless temperature after rescaling effects from minor embedding are taken into account by rescaling the unitless temperature by the chain strength.

\begin{figure}
\begin{centering}
    \includegraphics[width=7 cm]{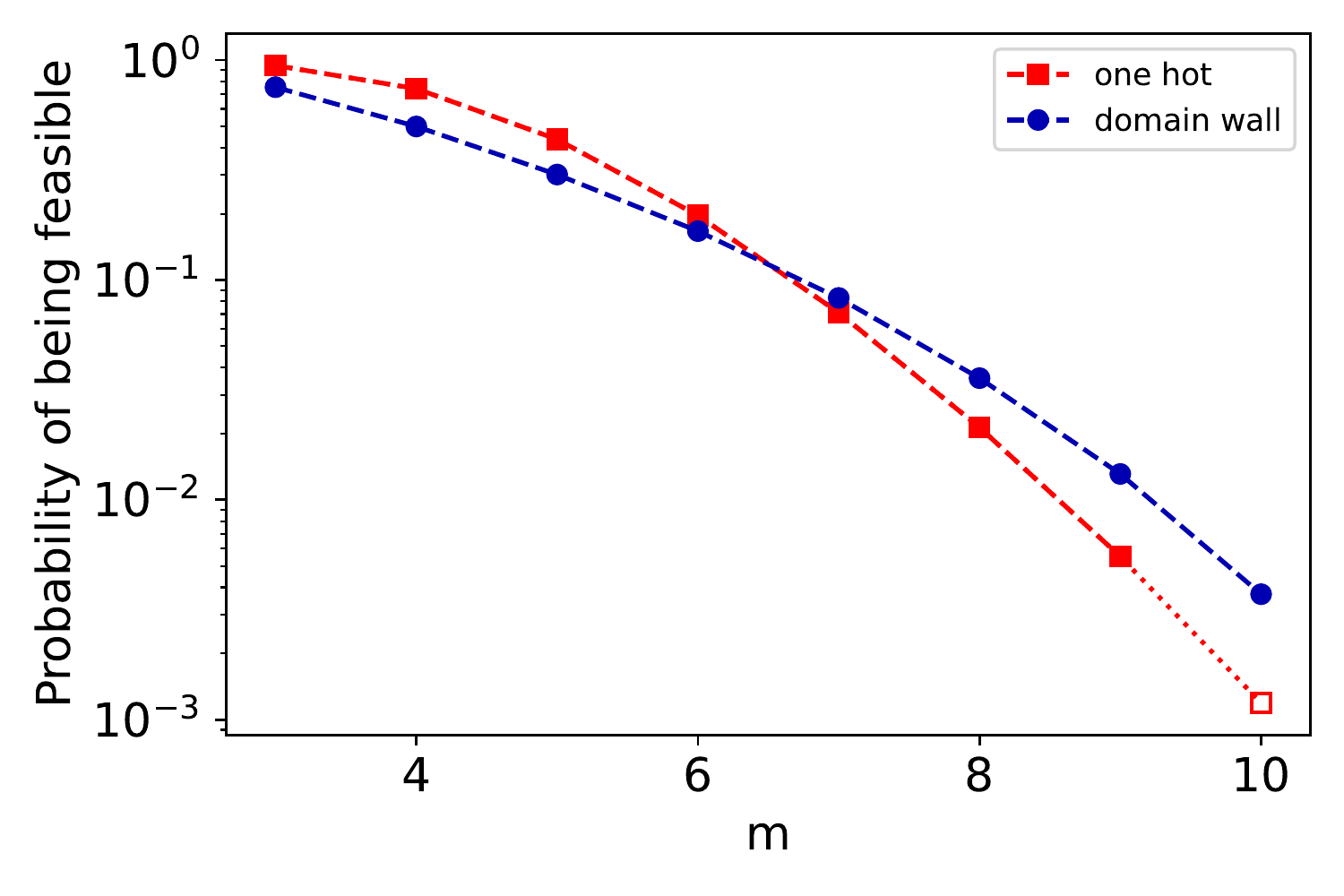}
    \caption{Probability to find feasible solution based on thermal equilibrium model at different values of $m$. This plot includes effects of minor embedding and assumes freezing at a fixed value of $B=5 \,GHz$ and a temperature of $15 \,mK \approx 0.31 \, GHz$. This plot is to illustrate the behaviour of the model, extracted values of the freeze time energy scales can be found in figure \ref{fig:oh_dom_E_scales}. All data points are based on $10^7$ samples, $95\%$ confidence statistical errorbars are smaller than depicted symbols.}
    \label{fig:thermal_success_prob_dw_oh}
    \par
\end{centering}
\end{figure}

Using the same Metropolis techniques as before, we can estimate the probability of feasible solutions at different values of $m$. The results are shown in Fig.~\ref{fig:thermal_success_prob_dw_oh}. Again, we see a crossover: For smaller $m$ the one-hot encoding performs better, but above $m=7$ the domain-wall method performs better. This is consistent with previous results, for smaller $m$ the effective temperature is smaller, and the difference in effective temperatures is greater. For larger $m$, on the other hand, the larger solution space is likely to have an increased effect given both the fact that the effective temperature is higher, and the ratio of the sizes of the solution spaces doubles each time $m$ is increased.

\subsection{Effective temperature\label{sub:eff_temp}}

In the previous section we examined the effect of encoding on success probability at thermal equilibrium. This thermal modelling can moreover be useful in evaluating the dynamics. In particular we can consider a model based on the KZM \cite{Kibble76a,Zurek96a,Damski05a,chancellor16b,Bando20a,Weinberg20a} where a system is approximated to remain in equilibrium until it approaches a phase transition and the dynamics suddenly freeze out. Furthermore, since it is the quantum fluctuations which mediate the dynamics, it is reasonable to assume that these have relatively little effect at the freezing point, and to a first approximation we consider the distribution at the freezing point to be a classical thermal distribution. This assumption may not be justified for all systems, particularly those which use ``free variable'' gadgets \cite{Boixo2013a,dickson13a,albash15a,chancellor16b,chancellor21a,chancellor21b}, where the transverse fields act at degenerate order in perturbation theory. It was further seen in \cite{chancellor16b}, that for problems which do not have ``free'' variables (all variables having having type 0 or I spin-sign transitions in the language of that work), the effects of transverse field and thermal fluctuations were very similar. 

Furthemore, all problems studied here are minor embedded, so therefore in many cases, multiple physical qubits must be flipped to change the value of a single logical variable, and the system must pass through high-energy ``broken chain'' states to do so. In other words, in a perturbative treatment similar to that used in \cite{Dodds19a}, the minimum perturbative order at which the transverse field can have an effect is equal to the minimum number of qubits in an embedding chain.  For one-hot encoding, all variables map to at least two qubits for $m>6$, while for domain-wall these variables map to at least two qubits for $m>7$. 

\begin{figure}
\begin{centering}
    \includegraphics[width=7 cm]{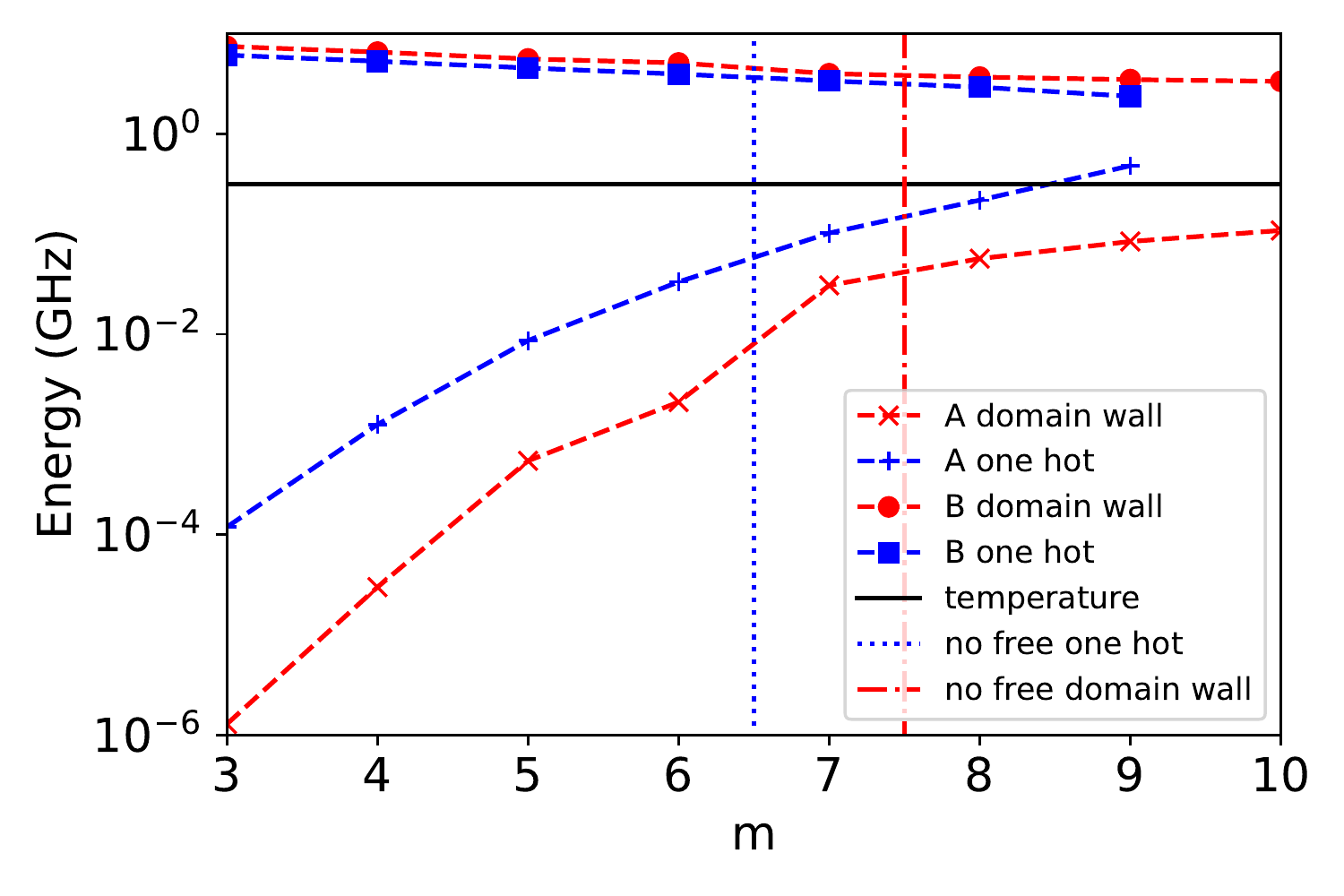}
    \caption{$A(s)$ and $B(s)$ energy scales at freeze time for the one-hot and domain-wall encoded unweighted assignment problem for different values of $m$. This is compared to the approximate temperature of $15 \, mK \approx 0.31 \, GHz$. Vertical lines (labeled ``no free one hot'' and ``no  free domain wall'') indicated the transition between values of $m$ where at least one variable in at least one embedding maps to a single qubit, and where every one maps to at least two qubits.}
    \label{fig:oh_dom_E_scales}
    \par
\end{centering}
\end{figure}

For lower values of $m$, we need to compare the strength of the quantum and thermal fluctuations at the freeze time, which can be estimated by comparing the experimental probability of feasible solutions to our model. If thermal fluctuations are much stronger than quantum fluctuations 
 then we are justified in our analysis. On the other hand, if it is the case that quantum fluctuations are on the order of thermal fluctuations (and some variables map to single qubits), then a model which explicitly includes quantum fluctuations is necessary. Fortunately, as we show later in figure \ref{fig:oh_dom_E_scales}, our classical treatment is justified. 

\begin{figure}
\begin{centering}
    \includegraphics[width=9cm]{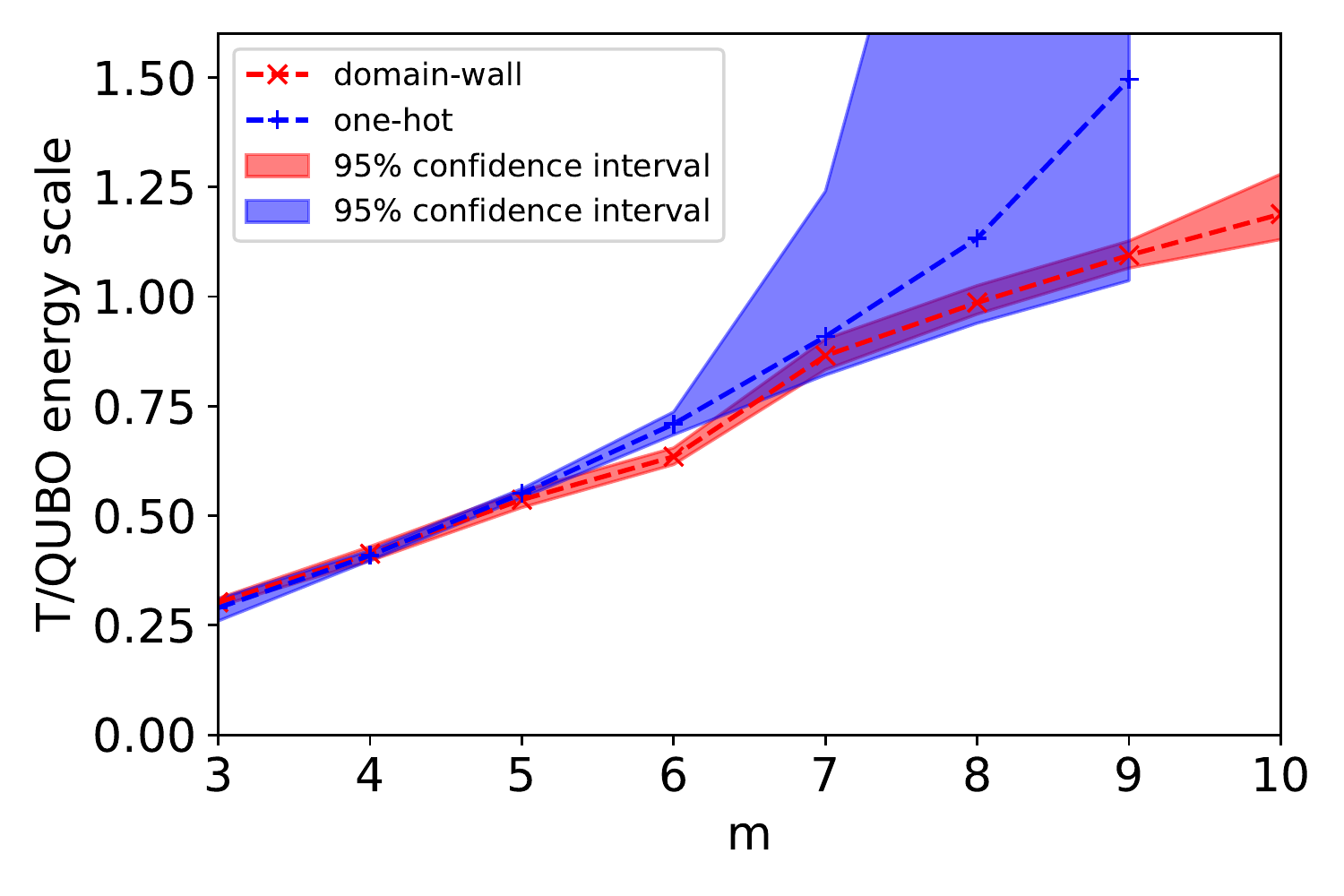}
    \caption{Effective temperature compared to QUBO energy scale at different values of $m$ for both encodings of the unweighted assignment problem. Methods for extracting the temperature are discussed in section \ref{sec:methods}.Note that the extremely large confidence intervals for one-hot at larger $m$ are primarily due to few feasible solutions being found, at the extreme of only a single feasible solution in all $10^5$ samples for $m=9$.} 
    \label{fig:T_QUBO}
    \par
\end{centering}
\end{figure}

To estimate the temperature we take the measured probability of a feasible solution and perform bisection until the model and the experiment match well. We first extract the unitless ratio of the temperature over the energy scale of the optimisation problem, ignoring for the moment the dynamic range squeezing caused by the strong ferromagnetic chains needed for minor embedding. The result is shown in Fig.~\ref{fig:T_QUBO}. From this figure we see little difference between the two encodings. 

\begin{figure}
\begin{centering}
    \includegraphics[width=7 cm]{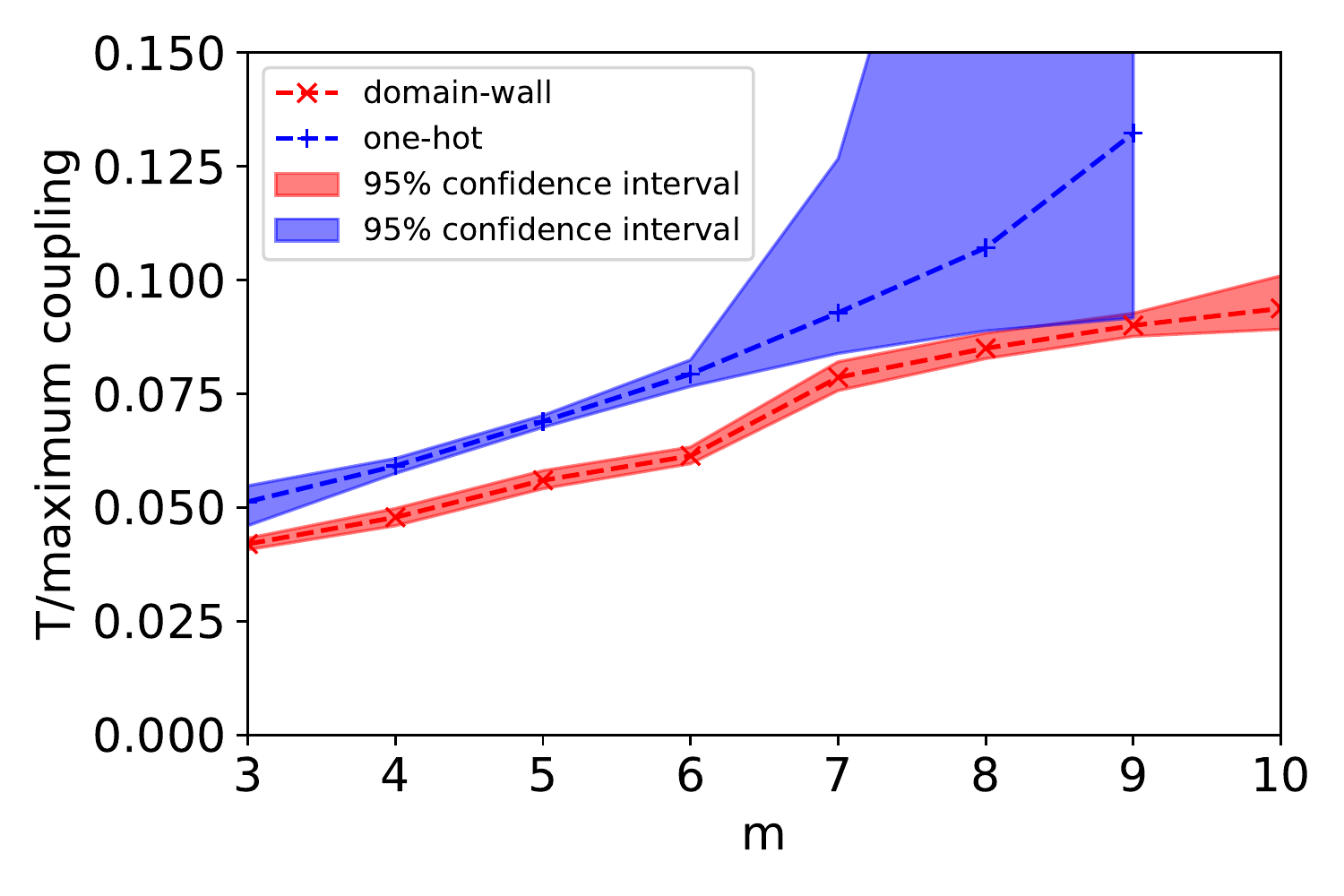}
    \caption{Effective temperature compared to maximum possible coupling at different values of $m$ for both encodings of the unweighted unweighted assignment problem. Methods for extracting the temperature are discussed in section \ref{sec:methods}.}
    \label{fig:T_coup}
    \par
\end{centering}
\end{figure}

\begin{figure}
\begin{centering}
    \includegraphics[width=7 cm]{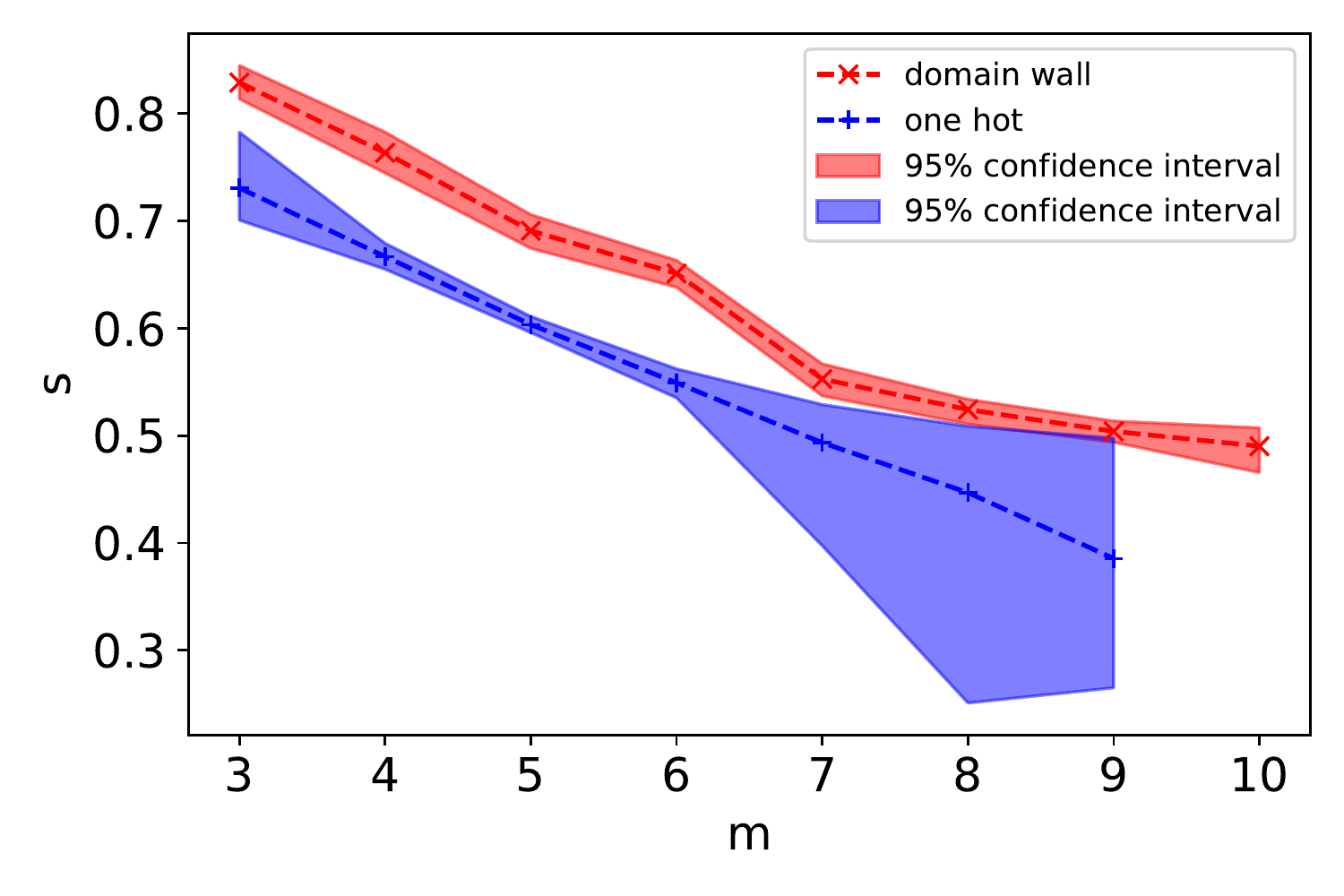}
    \caption{Extracted value of annealing parameter $s$ at the freezing point for one-hot and domain-wall encodings of the unweighted assignment problem at different sizes. Methods for extracting the value of $s$ are discussed in section \ref{sec:methods}.}
    \label{fig:oh_dom_s}
    \par
\end{centering}
\end{figure}

However, this figure does not tell the whole story, recall from Figs.~\ref{fig:uniform_torque_m_dw_oh} and \ref{fig:effective_temp_embed} that the heuristic we have used to choose the chain strength has specified a significantly weaker chain strength for one-hot. To really understand the freezing time we must compare the temperature not to the QUBO energy strength, but to the maximum coupling used for the problem. As Fig.~\ref{fig:T_coup} shows, once the overall energy scale is taken into account, it is clear that the domain-wall encoding is in fact sampling at a lower effective temperature. This result suggests that, the domain-wall encoding does indeed lead to later freezing of the dynamics (as depicted in figure \ref{fig:oh_dom_s}).

A later freezing time shows something fundamental about our encoding, namely that the dynamics of the annealing are facilitated by this encoding. The dynamics are more resistant to localization than they would be in one-hot. Since problem encoding is traditionally considered the domain of computer scientists, there is little (in fact no previous work we are aware of) research on how the physics is affected by how problems are encoded, this result shows that this is in fact an important consideration, and considering the encoding and device physics as completely separate ``layers'' of the device operation may not be an optimal approach.

We now return to the question of whether we were justified in ignoring quantum fluctuations in our final model of the distribution. To do this, we need to show two things, firstly we need to show that for regimes where some variables map only to single qubits the transverse field energy scale $A(s)$ is much lower than the temperature of the device, so that thermal fluctuations will have a stronger effect than quantum fluctuations. Secondly, in the regime where all variables are mapped to at least two qubits, we still need to show that $A(s)\ll B(s)$ so that treating the effect of the transverse field at second order in perturbation theory is justified. Again we can calculate these values by matching the success probability to an effective temperature in the thermal model and then extrapolating the energy scales based on an approximate temperature of $15 \,mK$. As Figure \ref{fig:oh_dom_E_scales} shows, these criteria are met for all problems studied here, suggesting that this model should at least be a reasonable approximation (although this approximation may begin to break down for one-hot at large $m$ values).

\section{Experimental and numerical Methods \label{sec:methods}}

In this section we explain the details of our experiments (both physical and numerical) for reproducibility purposes, for a more conceptual description of our tehcniques, see section \ref{sub:freezing}. All experiments reported here were performed on a D-Wave Advantage 1.1 quantum annealing computer which was accessed through Amazon Web Services between the dates of June 22 and June 25 2021. All experiments were performed using the default device settings, including an annealing time of $20 \,\mu s$. Spin reversal transforms (also known as gauge averaging) were not used. For each value of $m$, ten separate embeddings were used with $10,000$ anneals for each (for a total of $100,000$ anneals at each size). In addition to the sizes reported here, sizes $m=11$ and $m=12$ were also tested, but no feasible solutions were found with either encoding in all $100,000$ anneals. Furthermore, no experimental results are reported for one-hot with $m=10$ because no feasible solutions were found experimentally (for this reason we have left the $m=10$ symbol for one-hot unfilled on relevant theory plots). The annealing schedule and other device parameters were obtained from the D-Wave user forums \cite{forum-schedule}.

Effective temperatures for figures \ref{fig:T_QUBO} and \ref{fig:T_coup} were calculated by bisection using Monte Carlo. An initial temperature range was defined to run between $T_{\mathrm{min}}=0$ and $T_{\mathrm{max}}=2.5$ times the QUBO energy scale, then with $10^7$ samples at the midpoint of the range to test if the fraction of feasible solutions exceeded what was observed experimentally, if it did than a new $T_{\mathrm{min}}$ was defined to be this midpoint if not $T_{\mathrm{max}}$ was assigned the midpoint value. This routine was performed for $15$ iterations for each point. The $95\%$ confidence bounds were calculated using the same procedure, but using an experimental success probability which was the measured value plus or minus twice the experimental standard error. Quantities $A$, $B$ and $s$ were then calculated by first extracting $B$ based on a physical temperature of approximately $15\, mK\approx 0.31 \, GHz$, $A$ and $s$ were calculated by performing linear interpolation against an annealing schedule provided on the D-Wave user forums \cite{forum-schedule}. Note that these schedules are approximate and according to the source may deviate by up to $30\%$. Monte Carlo sampling was performed using standard Metropolis updates starting from a feasible solution. The number of Monte Carlo samples is based on the number of \textbf{attempted} updates (we have used $10^7$ to ensure statistical convergence and to ensure sufficiently many updates for the algorithm to achieve equilibrium, see appendix III), so cases where the update was attempted but not performed would still count as separate samples.

\section{Discussions and conclusions}

We have shown two important results related to the domain-wall encoding, and answered two key open questions. Firstly we have shown that for DQMs with general interactions, no better (quadratic) encoding can exist for problems with more than $3$ variables, which comprise all interesting optimisation problems for quantum computing applications. This result applies to any algorithm where a unary encoding of single variables is desired, this includes quantum annealing, but not exclusively, for example this method could equally well save qubits in gate model optimisation (in fact promising early results have been found \cite{plewa2021variational,Botelho2022mitigation}) and quantum inspired algorithms. Secondly, we have experimentally verified theoretical predictions that domain-wall encoding will lead to more favourable dynamics in physical implementations of quantum annealing. The first of these results is important because it implies that any search for more efficient (in terms of number of binary variables used) general DQM to QUBO encoding will be unsuccessful, and any future efforts can be focused on encoding specific structures of interactions or to settings where more general interactions are allowed and therefore our result does not hold. The second of these results highlights the importance of considering the physical dynamics the system will undergo when designing problem encodings. 

While problem encoding has traditionally be considered to be a computer science topic, our results suggest that there is important interplay between the encoding and the underlying physics, therefore the physics of the device needs to be considered when designing problem encodings.  While the problem statement used in the experiments reported here is not computationally hard, it is a common constraint structure for hard optimisation problems, such a the quadrtic assignment problem and travelling salesperson. Since the constraints will have to be stronger than the penalty terms which make the problems hare, they are likely to have a dominant effect on freeze time, therefore the results reported here are likely to carry over to those settings, at least qualitatively in the sense that domain-wall encoding leads to later freezing. On top of providing additional evidence that domain-wall encoding yields superior performance in the minor-embedded annealing setting, it also illuminates the mechanism, a fundamental change in the underlying dynamics of the system.

The goal of this study was not to fully understand how to optimise the performance of the device, so it is worth noting that other factors, such as anneal time will also change these properties. In particular, our results suggest that especially for larger QAP problems, annealing for longer to gain a later freeze time and therefore a lower effective temperature is likely to be fruitful. It is important to note that, while this work does provide useful relative comparisons of encodings, as with the work in \cite{chen21a}, it does not test the ultimate performance limits of the devices \footnote{potential routes to improvement are discussed in section III D of \cite{chen21a}}, so that the experiments remain simple. In particular the features which allow the minor embedding chains to effectively be twice as strong would probably improve the results substantially \cite{virtGraph} (this feature requires substantial calibration time so was not used in our experiment since we care about relative rather than absolute performance). 

While the experimental portion of this work was performed on a quantum annealing computer due to the larger sizes of devices available, other settings of particular interest are gate model settings, it is comparatively easier to engineer higher than quadratic interactions within this setting (typically at the cost of more circuit depth) leading to complicated tradeoffs \cite{Sawaya20a}, but it is likely that in at least some cases unary encoding of single variables are desired. There are also two-body drivers which drive directly between valid states for both one hot \cite{Hadfield19a} and domain wall \cite{chancellor2019}, but typically two qubit gates are noisier so this is again a tradeoff. In cases where unary encodings are used our work has a potentially strong impact, optimal QAOA \cite{Farhi14a} will mimic quantum annealing \cite{Brady21a} and therefore be subject to similar ``freeze time'' dynamics to those discussed here. The concept of a ``freeze time'' in fact affects any a system approaches a phase transition \cite{Kibble76a,Zurek96a} so may be relevant to other cases as well, for example in VQE \cite{Peruzzo14a} where the desired state lies near a phase transition.  Furthermore, coherent Ising machines \cite{McMahon16a,Inagaki16a} and digital annealers \cite{Yamaoka16a,Fujitsu_announce} both are limited to quadratic interactions and will be approximated by similar ``freeze time'' models to the ones we use here.

\section{Acknowledgments}

All authors were entirely supported in writing this paper by Quantum Computing Inc.~which also provided the necessary machine time.

\FloatBarrier
\section*{Appendix I: Travelling Salesperson as a DQM built on unweighted assignment constraints}

We start with the conventional statement of the travelling salesperson problem \cite{Bellman62a}: a salesperson must visit $m$ cities but can do so in any order and must choose the order which minimizes the total distance travelled, given distances $d_{\alpha,\beta}$ between each city. This problem may also include details such as requiring that the salesperson needs to start from and return to a base, and that each city can only be visited once \cite{Bellman62a}. We will first show a version without the base, and then show that additional linear terms can be added to represent the start and return to base. We also show that as long as the distances between the cities are positive, our mapping works even if a city can be visited multiple times. 

We consider a discrete quadratic model where the first discrete variable $x_1$ represents the choice of city to visit first, the second, $x_2$ represents the city visited second, the last represents the last, etc... Since there are $m$ cities the variables will have size $m$, and there will be $m$ of them. This constraint takes the form:
\begin{equation}
    \kappa\sum^{m-1}_{i,j=0}\sum^{m}_{\alpha=0}x_{i,\alpha}x_{j,\alpha},
\end{equation}
where $\kappa$ is a sufficiently strong positive weight to ensure the constraint is enforced. This constraint has the effect of incurring a cost if any of the two variables take the same values, in other words, if any city is visited twice. Furthermore since there are $m$ cities and $m$ variables, than no city being visited more than once implies that each is visited exactly once. To encode the distances between the cities, we add terms of the form $d_{\alpha,\beta}x_{i,\alpha} x_{i+1,\beta}$ which will add to the cost iff on a given leg of the journey, we start from city $\alpha$ and go to city $\beta$. The total problem statement becomes
\begin{equation}
    H_{\mathrm{TSP}}=\sum^{m-2}_{i=0}\sum^{m-1}_{\alpha,\beta=0}d_{\alpha,\beta}x_{i,\alpha} x_{i+1,\beta}+\kappa\sum^{m-1}_{i,j=0}\sum^{m}_{\alpha=0}x_{i,\alpha}x_{j,\alpha}.
\end{equation}
If we would like to further add a base from which the salesperson must start from and return to, we can add linear penalties to the first and last terms.
\begin{equation}
    H_{\mathrm{TSP+base}}=H_{\mathrm{TSP}}+\sum^m_{\alpha=0}d_{\mathrm{base},\alpha}(x_{0,\alpha}+ x_{m-1,\alpha}),
\end{equation}
where $d_{\mathrm{base},\alpha}$ is the distance between the base and city $\alpha$. 

One may now worry about what happens if we relax the condition that each city can only be visited once, in other words, what if the most efficient route to get to city $\beta$ from city $\alpha$ is to go through city $\gamma$ which may have already been visited. The answer is that as long as $d_{\alpha,\beta}$ represents this route where $d_{\alpha,\beta}=d_{\alpha,\gamma}+d_{\gamma\beta}$, the formalism still works, in other words, our representation makes no distinction between a route which involves passing through $\gamma$ if it has already been visited, or one which does not (assuming $\alpha$ is the $i$th city visited, this would be represented by setting $x_{i,\alpha}=1,\quad x_{i+1,\beta}=1$). Moreover if $\gamma$ hasn't already been visited than a route which \textbf{does} involve stopping at $\gamma$ can be represented (assuming $\alpha$ is the $i$th city visited) by setting $x_{i,\alpha}=1,\quad x_{i+1,\gamma}=1, \quad x_{i+2,\beta}=1$. As long as all $d_{\alpha,\beta}$ are positive the only way an optimal solution can involve re-visiting an already visited city is on the way to a new city, as discussed here, therefore our method of encoding the problem does support ``loops'' where the salesperson passes through the same city multiple times.

Our method does implicitly assume that the minimum distance between each pair of cities has already been calculated classically, however, finding the minimum distance between only two points is not computationally hard on a classical computer\cite{Dijkstra59a}. 

\section*{Appendix II: extensions of one-hot encoding}

\paragraph{k-hot constraints:}
A natural extension to the one-hot constraint is to consider the more general case where
\begin{equation}
    H_{\mathrm{k\mhyphen hot}}=\kappa \left( \sum_\alpha b_\alpha-\mathrm{k} \right)^2,
\end{equation}
where $\mathrm{k}$ is an integer between $1$ and $m-1$. These constraints only allow states where $\mathrm{k}$ of the binary variables take the $1$ value. Quadratic interactions must take the form $b_{i,\alpha}b_{j,\beta}$, and linear terms do not add independent degrees of freedom ($\sum_\beta b_{i,\alpha}b_{j,\beta}\propto b_{i,\alpha}+C$, where $C$ is an irrelevant offset). For a k-hot encoding where each discrete variable is encoded into $n_{\mathrm{var}}$ binary variables there will be $\binom{n_{\mathrm{var}}}{\mathrm{k}}$ possible configurations while quadratic and linear interactions only add $n_{\mathrm{var}}^2+2\,n_{\mathrm{var}}$ independent degrees of freedom while $\binom{n_{\mathrm{var}}}{\mathrm{k}}^2-1$ degrees of freedom are available (assuming each quadratic term within a variable is independent). It follows that (except for possibly in small instances) quadratic and linear terms can only realize general interactions for $k=1$ (one-hot) and $k=n_{\mathrm{var}}-1$ (one-hot with the definition of $0$ and $1$ reversed) where $\binom{n_{\mathrm{var}}}{\mathrm{k}}=\mathrm{k}$. These constraints are still very useful because the form of interactions between the discrete variables encode a structure where each of the binary variables taking a value of $1$ represents identical assets to be allocated, for example if  $\mathrm{k}$ identical pieces of equipment must be allocated over a total of $n_{\mathrm{var}}$ locations.
 
\paragraph{Integral encodings:}
Integral encodings can be thought of as products of k-hot encodings. Taking the specific form 
\begin{equation}
    H_{\mathrm{integral}}=\kappa \prod^R_q(\sum_\alpha b_\alpha-\mathrm{k}_q)^2
\end{equation}
where the coefficients $\mathrm{k}$ should be integers in the range $0$ to $m$, i.e.~$\mathrm{k} \in \mathcal{Z_R}$. The $R=1$ case is a $k\mhyphen hot$ encoding, but for all $R>1$ this constraint will require terms of order $2R$, which have to be realized by quadratization. Since k-hot encodings are not able to realize general interactions without higher than quadratic terms (assuming each state maps to a unique bitstring), integral encoding will not be able to either.

\section*{Appendix III: Convergence of Monte Carlo sampling}

\begin{figure}
\begin{centering}
    \includegraphics[width=7 cm]{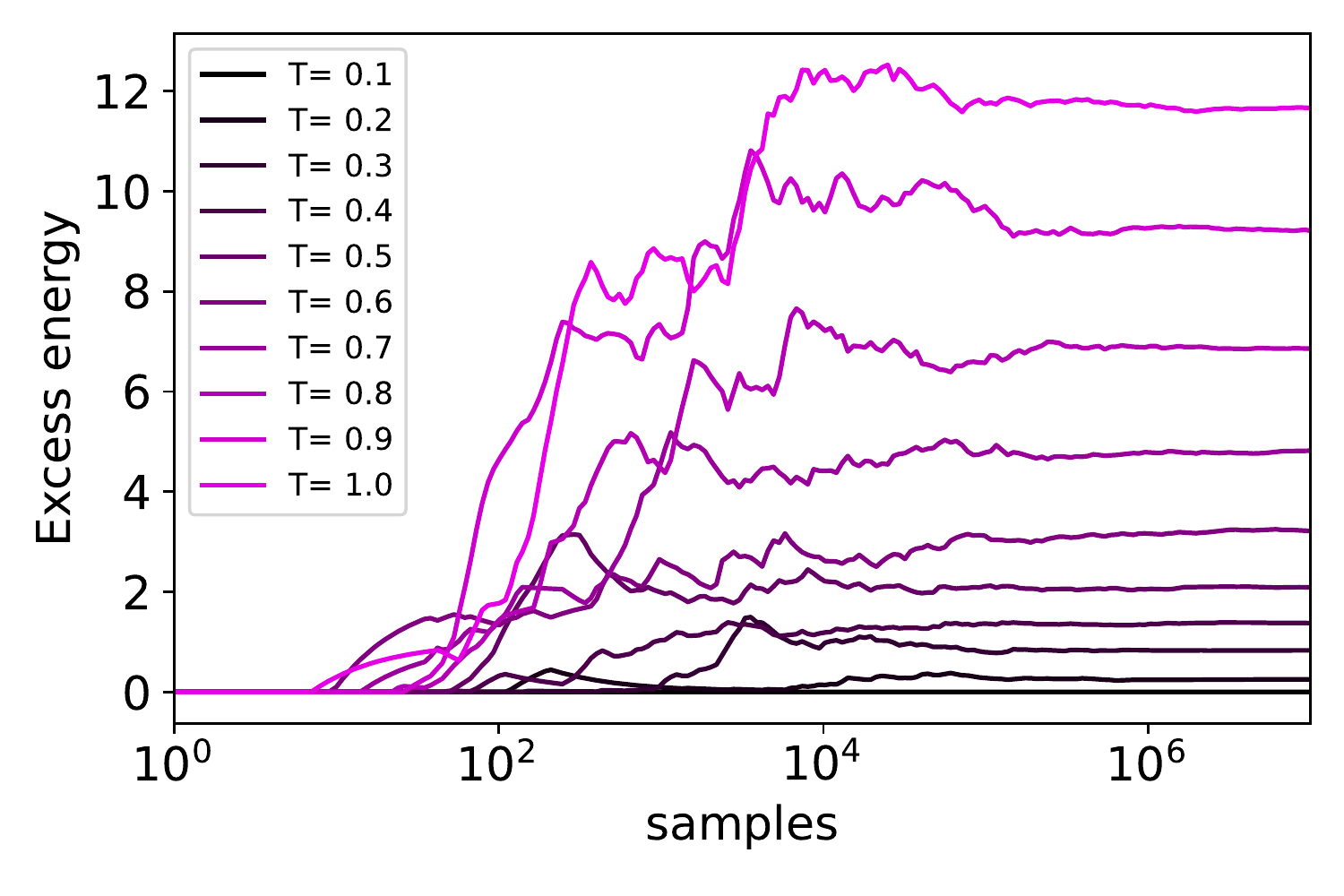}
    \caption{Running average of excess energy (compared to a feasible state) for an unweighted $m=10$ assignment problem with up to $10^7$ samples at different dimensionless temperatures encoded using the domain-wall encoding. Note logorithmic scale of x-axis.}
    \label{fig:E_conv_dom_n=10}
    \par
\end{centering}
\end{figure}

\begin{figure}
\begin{centering}
    \includegraphics[width=7 cm]{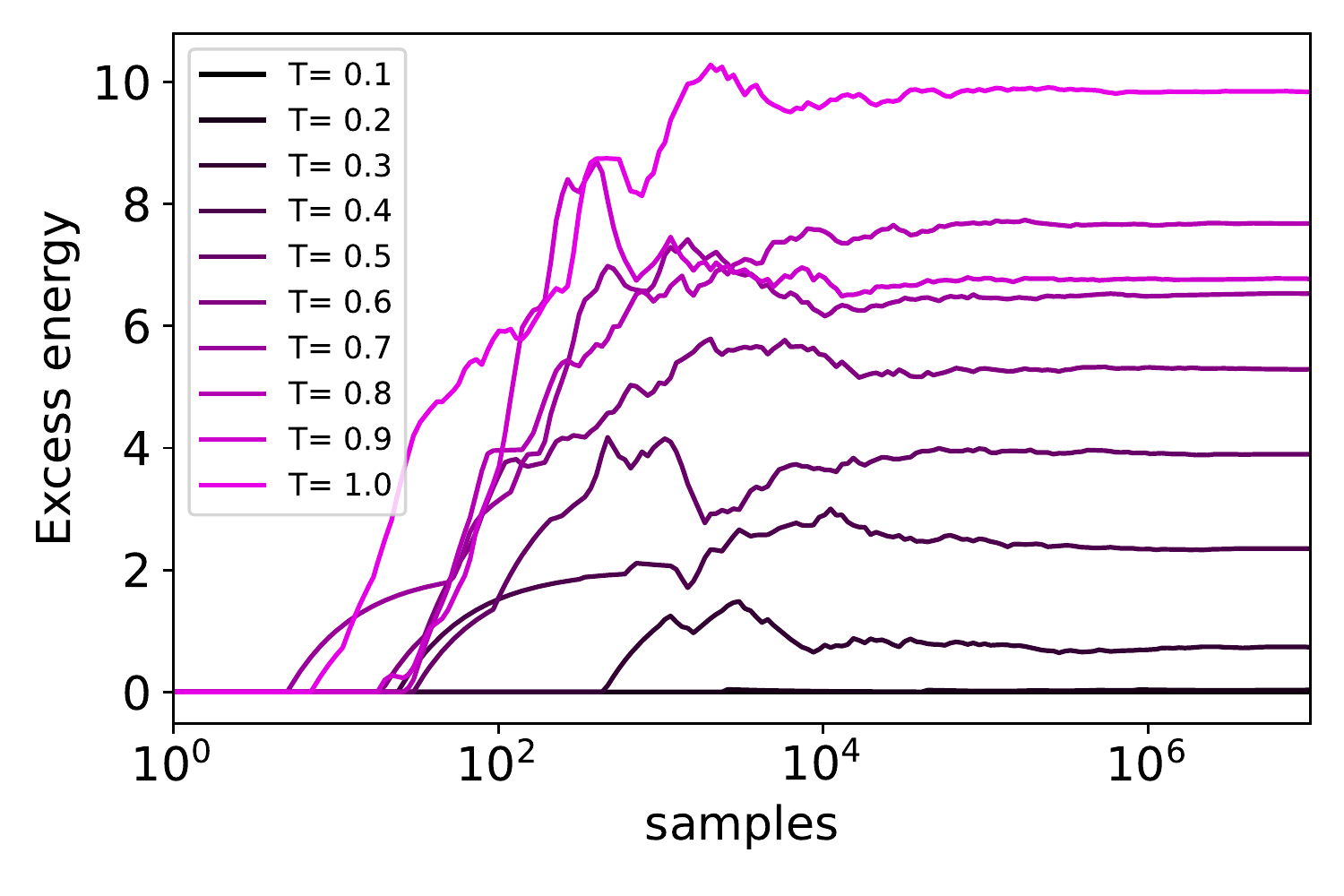}
    \caption{Running average of excess energy (compared to a feasible state) for an unweighted $m=10$ assignment problem with up to $10^7$ samples at different dimensionless temperatures, encoding using one-hot encoding. Note logorithmic scale of x-axis.}
    \label{fig:E_conv_oh_n=10}
    \par
\end{centering}
\end{figure}

The experimental data analysis in this work relies on the numerical Monte Carlo sampling being able to accurately sample a thermal distribution. While the high level of symmetry of the unweighted assignment problem suggests that equilibration should be relatively quick, we should still verify this numerically. As we can see from figures \ref{fig:E_conv_dom_n=10} and \ref{fig:E_conv_oh_n=10}, even for the largest system here, complete convergence has occurred after $10^6$ samples at all temperatures, and a fair quality approximate convergence has occurred even by $10^5$ samples.

While these plots show convergence, the remaining question is rather subsequent runs will typically reach a similar equilibrium. To check this we perform for each temperature and both encodings ten times with $10^7$ samples each. We than find the ratio of the standard deviation divided by the mean over each set of ten samples. Standard deviation is used rather than standard error, since many of our methods rely on a single set of $10^7$ samples. What we find is that for most cases the standard error relative to mean is small with a few notable exceptions. The sampling is not reliable below $\frac{T}{\mathrm{QUBO \,energy}}=0.2$ for either encoding, the domain-wall encoding shows a large normalized standard deviation at $\frac{T}{\mathrm{QUBO \,energy}}=0.1$, while the ratio is  not defined for the one-hot encoding because in some cases no unfeasible solutions were found. Fortunately this is not very relevant to our experimental analysis since the lowest experimentally fitted value of unitless temperature is around $\frac{T}{\mathrm{QUBO \,energy}} \approx 0.25$ (see figure \ref{fig:T_QUBO}). It is likely that this relatively large normalized standard deviation is due to the fact the the mean energy is very small. The normalized standard deviation for $\frac{T}{\mathrm{QUBO \,energy}}=1.0$ for domain wall is also relatively large, and might somewhat affect the accuracy of the extracted temperature for $m=10$, but this point cannot change our key conclusions because there is no one-hot data to compare with. We have not checked temperatures above $\frac{T}{\mathrm{QUBO \,energy}}=1.0$ but the only experimental value which significantly exceeds this temperature is the one-hot sample at $m=9$, since this point already has a large experimental standard error, it is not relevant to check that the contribution to the error from sampling is significant.

\begin{table}
\begin{tabular}{|c|c|c|}
\hline 
T & domain wall & one-hot\tabularnewline
\hline 
\hline 
0.1 & 0.17 & N/A\tabularnewline
\hline 
0.2 & 0.016 & 0.039\tabularnewline
\hline 
0.3 & 0.0063 & 0.013\tabularnewline
\hline 
0.4 & 0.0030 & 0.004\tabularnewline
\hline 
0.5 & 0.0069 & 0.0028\tabularnewline
\hline 
0.6 & 0.095 & 0.0012 \tabularnewline
\hline 
0.7 & 0.0032 & 0.095\tabularnewline
\hline 
0.8 & 0.046 & 0.080 \tabularnewline
\hline 
0.9 & 0.00316 & 0.0013 \tabularnewline
\hline 
1.0 & 0.11 & 0.062\tabularnewline
\hline 
\end{tabular}

\caption{\label{tab:rel_std} Standard deviation (NB \textbf{not} standard error) over mean energy value for $10$ Monte Carlo runs with $10^7$ samples each at different temperatures for both encodings.}

\end{table}

\FloatBarrier

\bibliography{reference}

\end{document}